\documentclass[11pt]{article}
\pdfoutput=1
\usepackage{graphicx}
\usepackage{latexsym}
\usepackage{amsmath}
\usepackage{amsfonts}
\usepackage{mathrsfs}
\usepackage{dsfont}
\usepackage{enumerate}
\usepackage{verbatim}
\usepackage{hyperref}
\usepackage{tikz}
\usepackage{mathtools}
\usepackage{simplewick}
\usepackage{float}
\usepackage{caption}
\usepackage{slashed}
\usepackage[title]{appendix}
\usepackage{graphicx}
\usepackage{subfig}

 \csname
@addtoreset\endcsname{equation}{section}
\usepackage{amsmath,amssymb}
\usepackage{graphicx}
\usepackage{color}

\begin{document}

\def\be{\begin{equation}}
\def\ee{\end{equation}}
\def\bea{\begin{eqnarray}}
\def\eea{\end{eqnarray}}
\def\nn{\nonumber}

\renewcommand{\thefootnote}{\fnsymbol{footnote}}
\renewcommand*{\thefootnote}{\fnsymbol{footnote}}

\begin{flushright}

\end{flushright}

\vspace{40pt}

\begin{center}

{\Large\sc One-loop corrections to the primordial tensor spectrum from massless isocurvature fields}

\vspace{50pt}

{\sc Hai Siong Tan}

\vspace{15pt}
{\sl\small Division of Physics and Applied Physics,
School of Physical and Mathematical Sciences, \\
Nanyang Technological University,\\
21 Nanyang Link, Singapore 637371\\[5ex]
Physics Division, National Center for Theoretical Sciences,\\
Hsinchu 30013, Taiwan. }

\vspace{15pt}

\vspace{70pt} {\sc\large Abstract}\end{center}

We study one-loop corrections to the two-point correlation function of tensor perturbations
in primordial cosmology induced by massless spectator matter fields.
Using the Schwinger-Keldysh formalism in cosmological perturbation theory, 
we employ dimensional regularization and cutoff regularization to study the finite quantum corrections  at one-loop 
arising from isocurvature fields of 
the massless scalar, fermion and abelian gauge field which are freely propagating on the FRW spacetime. 
For all cases, we find a logarithmic running of the form
$\frac{C}{q^3} \, \frac{H^4}{M^4_p}
 \log \left( \frac{H}{\mu } \right)$ where $C$ is a negative constant related to the beta function, $H$ is the Hubble parameter at horizon exit 
and $\mu$ is the renormalization scale.

\newpage

\tableofcontents

\renewcommand*{\thefootnote}{\arabic{footnote}}
\setcounter{footnote}{0}

\section{Introduction}
\label{sec:Introduction}

The Schwinger-Keldysh formalism has been the underlying
quantum field theoretic framework for computing correlation functions in cosmological perturbation theory, many
crucial aspects of which were presented in the seminal paper by Weinberg in \cite{Weinberg}.
Essential cosmological observables such as the primordial spectra and bispectra are derived within 
the validity of this
formalism which carries with it the prescription to compute loop effects for these correlation functions. 
In this paper, we will employ this formalism in the context of inflation
to compute quantum corrections at one-loop to the primordial tensor spectrum 
which at tree-level reads, in momentum space and up to slow-roll corrections,
\be
\label{treeTensor}
\Delta^2_t (k) = \frac{2}{\pi^2 k^3} \left( \frac{H (\tau_k) }{M_p} \right)^2 ,
\ee
with $H (\tau_k ) = k/ a(\tau_k )$ is the Hubble scale at horizon crossing. 
The loop corrections that we study in this paper pertain to those arising from massless spectator/isocurvature
fields of spin $<2$ freely propagating on the cosmological spacetime, and
we compute the various corrections to \eqref{treeTensor} with the slow-roll approximation being invoked in our expressions for the graviton mode and scalar wavefunctions in the comoving gauge. 

In the context of primordial inflation, loop corrections to the inflaton two-point function have been considered in several works after the appearance of \cite{Weinberg}.
Among the most principal results is that of Senatore and Zaldarriaga in \cite{Senatore} where
they showed how the one-loop correction to the scalar spectrum is consistent with scale invariance
contains a logarithmic dependence on the renormalization scale as follows
\be
\label{scalarSenatore}
\langle \zeta^2_k \rangle_{1-loop} \sim \frac{\beta}{k^3} \log \left( 
\frac{H (\tau_k )}{\mu} \right),
\ee
where $\mu$ is the renormalization scale
and $\beta$ is a constant
that depends on the type of matter field that couples to the inflaton - 
this was computed in \cite{Weinberg}, \cite{Chai} and \cite{Tan} for the cases of minimally and conformally coupled
scalar fields, Dirac fermion and abelian gauge field. The result \eqref{scalarSenatore} is valid at leading order of the slow-roll expansion and contributes to the fully renormalized correlation function obtained after taking into account the complete renormalization
prescription of the theory.

Although there has been a sizable amount of literature
discussing loop corrections to cosmological correlation functions (see for example \cite{Tanaka} for a nice review), 
much less attention has been devoted specifically to loop corrections to the tensor spectrum. 
In \cite{Feng}, a generalization of \cite{Chai} was done to compute quantum correction at one-loop 
level to the tensor spectrum due to free massive and massless Dirac fermions and it led to a result similar to 
\eqref{scalarSenatore} with $\beta$ being more technically involved to compute and smaller by a factor of 
$\frac{H^2}{M^2_p}$. There was also a similar attempt to compute the one-loop correction due to scalars in \cite{ Adshead, Eugene}. 

An appropriate regularization procedure for making sense of the loop integrals is
required to extract the finite one-loop correction, and we found that the methods first presented in \cite{Weinberg,Senatore}
were not quite accurately implemented in the papers discussing loop corrections to the scalar and tensor spectra. 
For example, in an earlier work in \cite{Tan}, we revisited the calculation
for the scalar spectrum and corrected the factors of $\beta$ in \eqref{scalarSenatore} as
presented in \cite{Chai}. Loop corrections also arise in the computation of
potentially discoverable observables like non-gaussianity and in particular they can harbor signatures of matter fields dynamically present during inflation. Such calculations rely on the correct implementation of the Schwinger-Keldysh formalism and thus there is good motivation
for understanding how to regularize divergent loop integrals correctly. In this paper,  we will use both dimensional regularization and cutoff regularization. For the former, we furnish some details (crucial for concrete computations) missing (or implied) in both \cite{Weinberg} and \cite{Senatore} which can be relevant for other loop computations within the general context of cosmological perturbation theory. 
This was first done for a related context ( loop corrections to the scalar spectrum ) in \cite{Tan}.

In this paper, we consider massless isocurvature fields minimally coupled to the FRW background, specifically the real scalar, Dirac fermions, abelian gauge field.
We find that , the one-loop corrections due to fields of various spins contain logarithmic terms 
of the following form in momentum space
\be
\label{tensoroneloop1}
\langle
h_{mn} h_{mn} \rangle_{1-loop} = \frac{C}{q^3} \frac{H^4}{M^4_p} \log \left( \frac{H}{\mu} \right), \,\,\, H = \frac{q}{a(\tau_q)},
\ee
where $h_{mn}$ is the tensor perturbation gauge-fixed to be transverse and traceless, and
$C$ is a negative constant. This contributes to the finite one-loop correction which is obtained in principle
after a full renormalization is implemented, an issue which we would not pursue here but would be interesting as a follow-up. 
At tree-level, it was shown in \cite{Mark2017} that the renormalization of 
\eqref{treeTensor} yields finite correction terms that 
vanishes in the zero $k$ limit. The technique employed in \cite{Mark2017}
 was that of adiabatic subtraction \cite{Mark2013}, which for free isocurvature fields propagating on FRW spacetime, 
amounts to a redefinition of the bare cosmological constant and Planck mass via counter-terms\footnote{See also \cite{Senatore} for a discussion of counter-terms for \eqref{scalarSenatore} in an effective field theory formulation}.  
It would be relevant to study how the renormalization prescription in \cite{Mark2013,Mark2017}
extends to the tensor spectrum at one-loop level where our result \eqref{tensoroneloop1} should play a crucial role.

The outline of our paper is as follows: in Section~\ref{sec:General}, we provide the general scheme of the one-loop computation including
the two regularization methods and some details concerning the graviton 4-point function which are universally relevant for all the matter fields considered. This is followed by Section~\ref{sec:Matter} where we present some technical details specific to each type of matter field
and the one-loop correction constant $C$ in \eqref{tensoroneloop1} for each. In Section~\ref{sec:Seagull}, we briefly explain why all seagull vertices in the interaction Hamiltonians do not contribute to the one-loop logarithmic term in \eqref{tensoroneloop1} for all the cases considered. Finally, in Section~\ref{sec:Discussion}, we conclude with a summary of results and outlook.

\section{General aspects of the one-loop computation}
\label{sec:General}

\subsection{Schwinger-Keldysh formalism}
\label{sec:SKsection}

In the Schwinger-Keldysh (or `in-in') formalism\footnote{See for example \cite{Liam} and \cite{TASI} for a good review of this topic.}, we compute the two-point correlation function of 
the tensor perturbation $\langle h_{mn} (\vec{x}, \tau ) h_{mn} (\vec{x}', \tau ) \rangle$ 
evaluated at some common late time $\tau$ in the interaction picture, with the prescription 
\bea
\label{SK}
&&\langle \Omega | h_{mn} (\vec{x}, \tau ) h_{mn} (\vec{x}', \tau ) | \Omega \rangle \cr
& =& 
\langle 0 | \left[  
\bar{T} \text{exp} \left( i \int^\tau_{-\infty_+} dt\, H_{int} (t) 
\right) 
\right] h_{mn} (\vec{x}, \tau ) h_{mn} (\vec{x}', \tau )
\left[  
T \text{exp} \left( - i \int^\tau_{-\infty_-} dt\, H_{int} (t) 
\right) 
\right] |0 \rangle, \nonumber \\
\eea
where the interaction Hamiltonian $H_{int} (t)$ is obtained by
first carrying out a perturbation of the background metric in the Lagrangian, collecting all terms up to quadratic order and then performing the standard Legendre transform of each dynamical field. Henceforth, we take $\tau \approx 0$. 

\begin{figure}[h]
  \centering
  \subfloat[$\,$]{\includegraphics[width=0.4\textwidth]{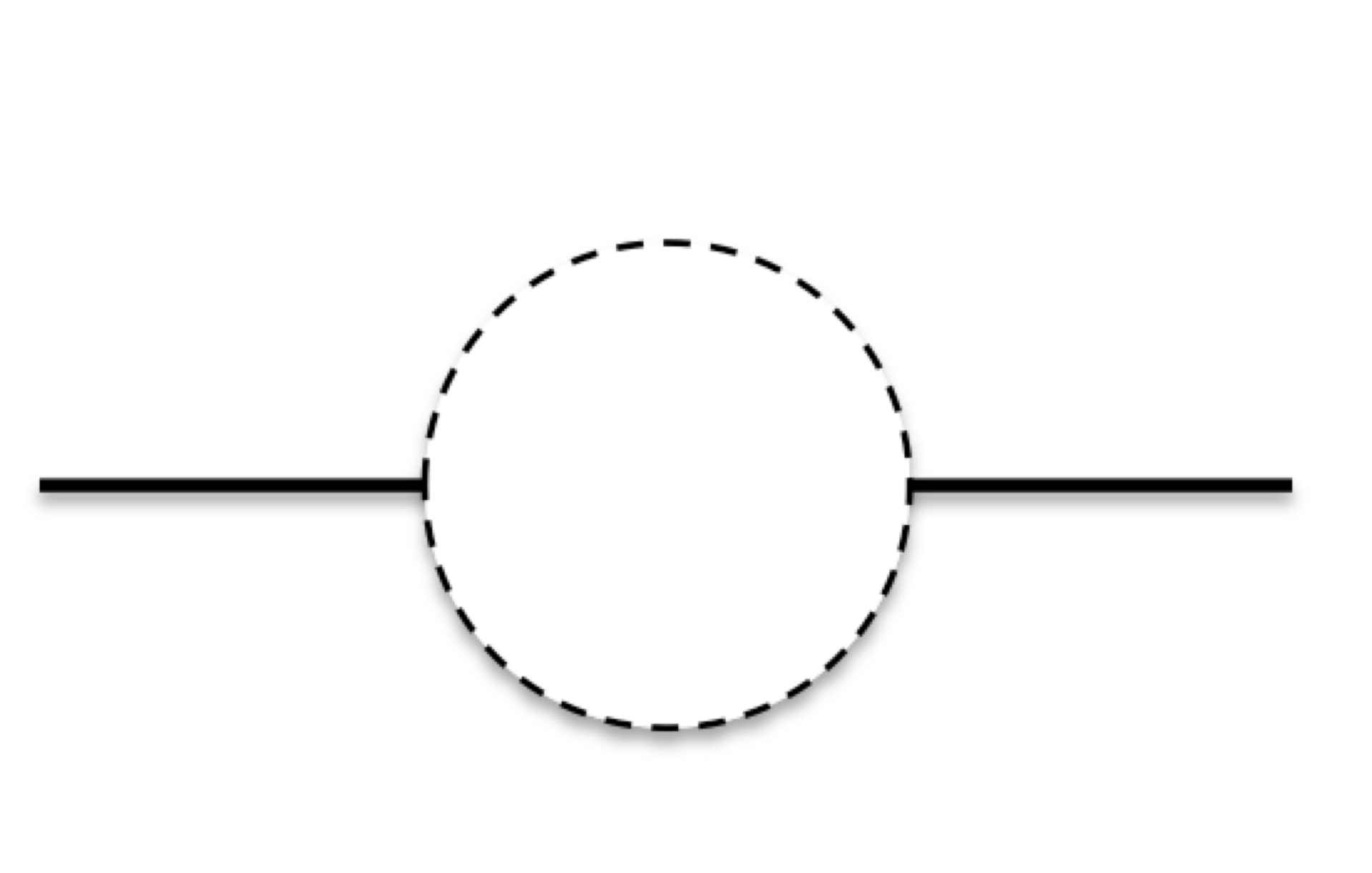}\label{fig:f1}}
  \hfill
  \subfloat[$\,$]{\includegraphics[width=0.4\textwidth]{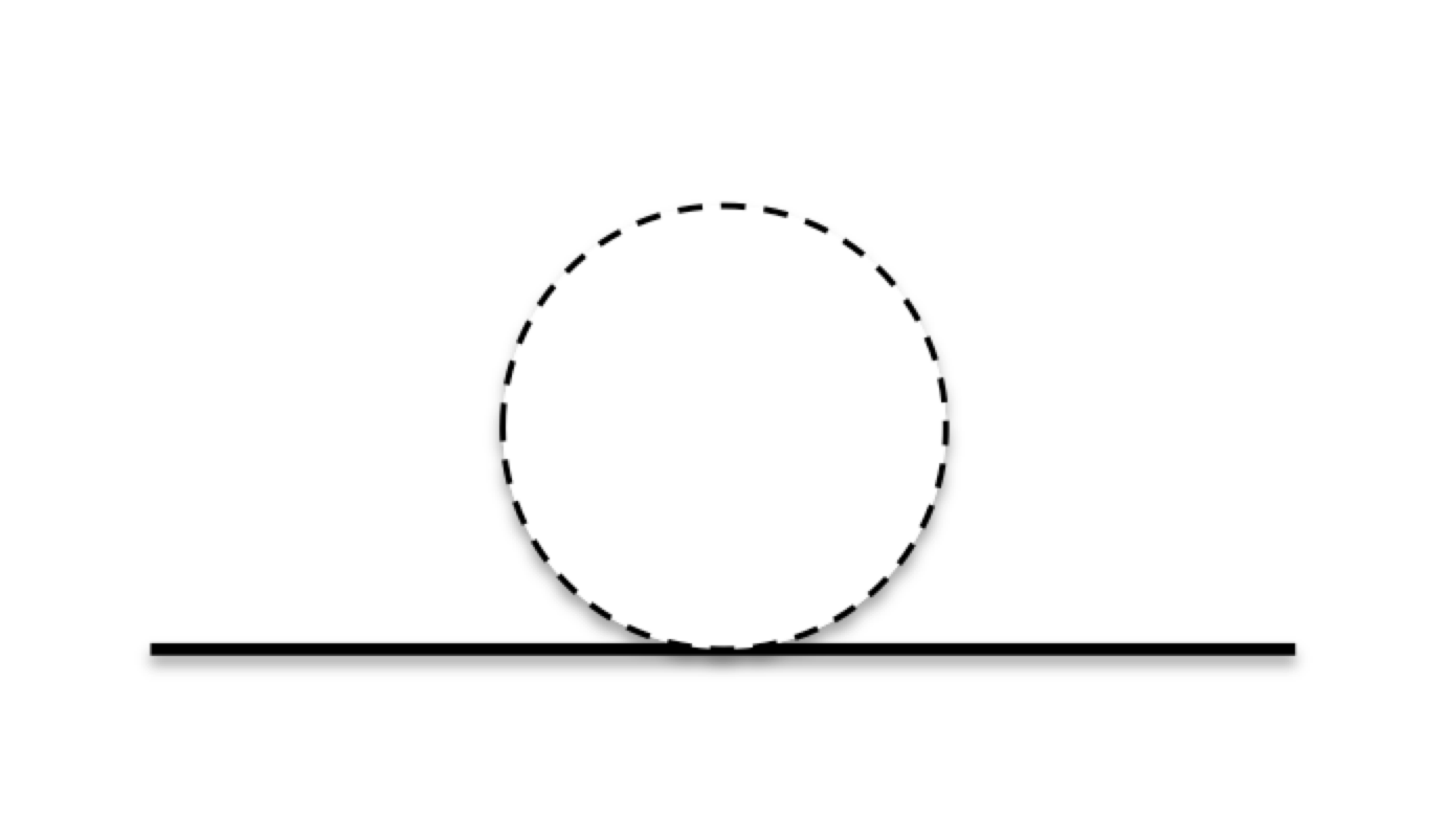}\label{fig:f2}}
  \caption{For the matter fields we consider in this work, 
all interaction Hamiltonians that give rise to the one-loop correction term are of the form diagrammatically 
represented by (a) whereas `seagull' vertices of the form (b) do not contribute as we explain in Section~\ref{sec:Seagull} .}
\end{figure}

In \eqref{SK}, the vacuum state of the free theory $| 0 \rangle$ is obtained after projecting on the interacting vacuum state $| \Omega \rangle$ with an $i\epsilon$ presciption, with the infinities analytically continued as 
\be
\infty_{\pm} = \infty (1 \pm i\epsilon ),
\ee
with $\epsilon$ being a real and positive regulator. In the interaction picture, the field operators evolve via the free Hamiltonian and thus they are expanded in terms of modes which are solutions to the Mukhanov free field equations. The vacuum expectation value is then computed in perturbation theory by expanding \eqref{SK} to the required order. In this work, we focus on one-loop corrections to the primordial spectrum which, for the theory that we consider - Einstein gravity coupled to various matter fields - arise from terms in \eqref{SK} up to and including the following second-order terms 
\bea
\langle h_{mn} (\vec{x} ) h_{mn} (\vec{x'} ) \rangle &=& 
-2 \text{Re} \left(   
\int^0_{-\infty_+} d\tau_2 \int^{\tau_2}_{-\infty_+} d\tau_1 \langle 0 | H_1 H_2 h_{mn} (\vec{x} ) h_{mn} (\vec{x'} )
 |0 \rangle
\right)\cr
\label{SK1}
&&\qquad \qquad
+ \int^0_{-\infty_+} d\tau_1 \int^0_{-\infty_-} d\tau_2 \langle 0 | H_1  h_{mn} (\vec{x} ) h_{mn} (\vec{x'} )
 H_2 | 0 \rangle
\eea
where $H_{1,2} = \int d^3x_{1,2} \mathcal{H}_{int} (\tau_{1,2}, \vec{x}_{1,2} )$. 
As we shall see later in Section~\ref{sec:Seagull}, the first-order terms of the form 
$$ 
- 2 \text{Im} \left(  \int^0_{-\infty_+} d\tau \langle H_{int} (\tau) h_{mn} (\vec{x},0) h_{mn} (\vec{x}',0)\rangle \right)
$$
do not contribute to the one-loop logarithimic correction term for the matter fields that we consider. 
Diagrammatically, we can represent the second- and first-order terms in Figure \ref{fig:f1} and \ref{fig:f2} respectively.

We work in the spatially flat gauge, with the tensor perturbations defined as 
\bea
\,\,\,
g_{ij} &=& b_{ij} + \gamma_{ij}, \,
b_{ij} = a^2 \delta_{ij}, \,\, \gamma_{ij} = a^2 h_{ij},
 \cr
g^{ij} &=& b^{ij} - \gamma^{ij} + {\gamma^i}_k \gamma^{kj} + \ldots, \,
\eea
with $h_{ij}$ being transverse and traceless. In this paper, we work in the conformal chart of the FRW geometry
with vierbein $e^\mu_a = \frac{1}{a(\tau )} \delta^\mu_a$. Relative to the primordial scalar spectrum, the
tensor spectrum is smaller by a factor of the slow-roll parameter $\epsilon$ and in this paper, we will keep to the lowest order in the slow-roll expansion. In particular, this implies that in computing the one-loop correction, we will invoke the approximation of adopting the graviton and scalar mode wavefunctions to be those on planar de Sitter spacetime at the time of horizon exit when integrating over the internal loop. For the photon and massless fermion wavefunctions, this is not required as their classical equations of motion are conformally invariant.

In evaluating the correlation functions, we note that the terms which 
contribute to the one-loop quantum correction arise from terms in $H_{int}$
which are linear in $h_{mn}$. Thus, in \eqref{SK1}, each correlation function factorizes into 
a product of a 4-point function of matter fields (and their derivatives) and a
4-point function of the graviton field $h_{mn}$. 
There are two time integral contours in \eqref{SK1} which contain the same matter field
4-point function but slightly different graviton 4-point functions. The main 
technicality involved in the computation lies in the simplification 
of the index contractions between these two 4-point functions 
which will be elaborated in the subsequent sections.

\subsection{Graviton 4-point function}
\label{sec:Graviton}

The graviton field can be expanded in terms of bosonic oscillators $\hat{a}_{\vec{q}, \lambda }$, mode wavefunctions 
$h_q (\tau )$ and rank-two polarization tensors  $\epsilon_{ij}$ as follows ($q \equiv |\vec{q} |$)
\be
\label{gravitonmodes}
h_{ij} = \int d^3q \sum_\lambda e^{i\vec{q} \cdot \vec{x}} \epsilon_{ij} (\hat{q}, \lambda ) \hat{a}_{\vec{q}, \lambda } h_q (\tau )
+ c.c.
\ee
where the modes satisfy the Mukhanov equation
\be
V''_k + \left( k^2 - \frac{a''}{a} \right) V_k = 0, \qquad V_k (\tau) = a (\tau) M_p h_k (\tau),
\ee
where we have used primes to denote derivative with respect to conformal time $\tau$. 
Given the full inflationary background equations that determine $a(\tau )$, one can then solve for the modes $h_k (\tau)$. In the context of slow-roll inflation, we take the spacetime near horizon-crossing to be quasi-de Sitter 
and up to slow-roll corrections, the mode wavefunctions read \footnote{To connect our normalization conventions to that usually stated for the tree-level primordial tensor spectrum $\Delta^2_t$, simply multiply the 2-point function (of the graviton fields) obtained from \eqref{gravitonmodes} by $1/(2\pi )^2$.}
\be
h_q (\tau) = \frac{\sqrt{16\pi G}}{(2\pi )^{3/2}} \frac{H (\tau_q) }{q^{3/2}} (1+iq \tau ) e^{-iq \tau} + \mathcal{O}(\epsilon ).
\ee
The bosonic oscillators obey 
$$
[ \hat{a}_{\vec{q}, \lambda }, \hat{a}_{\vec{q}', \lambda' } ]  = \delta^3 ( \vec{q} - \vec{q}' ) \delta_{\lambda \lambda'},
$$
and the polarization tensors are normalized in this paper as (see for example \cite{Weinberg:text})
\be
\sum_\lambda \epsilon_{ij} (\hat{q}, \lambda )   \epsilon^*_{kl} (\hat{q}, \lambda ) = \delta_{ik} \delta_{jl} + \delta_{il} \delta_{jk} - \delta_{ij} \delta_{kl} 
+ \delta_{ij} \hat{q}_k \hat{q}_l + \delta_{kl} \hat{q}_i \hat{q}_j - \delta_{ik} \hat{q}_j \hat{q}_l - \delta_{il} \hat{q}_j \hat{q}_k - \delta_{jk} \hat{q}_i \hat{q}_l - \delta_{jl} \hat{q}_i \hat{q}_k  + \hat{q}_i \hat{q}_j \hat{q}_k \hat{q}_k,
\ee
\bea
\sum_{\lambda, \lambda'} \epsilon_{kl} (\hat{q}, \lambda) \epsilon^*_{mn} (\hat{q}, \lambda ) \epsilon^*_{ij} 
(\hat{q}, \lambda' ) \epsilon_{mn} (\hat{q}, \lambda' ) 
\label{tensorp2}
&=& 2 \left( \delta_{ki} \delta_{lj} - \delta_{ki} \hat{q}_l \hat{q}_j - \delta_{lj} \hat{q}_k \hat{q}_i + \hat{q}_k
\hat{q}_l \hat{q}_i \hat{q}_j \right)\cr
&&\qquad+ 
2 \left[
\delta_{kj} \delta_{li} - \delta_{kl} \delta_{ij} + \delta_{ij} \hat{q}_k \hat{q}_l - 
\delta_{li} \hat{q}_k \hat{q}_j + \delta_{kl} \hat{q}_i \hat{q}_j - \delta_{kj} \hat{q}_l \hat{q}_i 
\right]. \nonumber  \\
\eea
For subsequent calculations, it is rather useful to note the following properties : 
(i)$\epsilon^*_{mn} (-\vec{q}, \lambda ) = \epsilon_{mn} (\vec{q}, \lambda )$, 
(ii)\eqref{tensorp2} is invariant under the exchanges 
$i\leftrightarrow j$, $k\leftrightarrow l$ and 
$\{ k \leftrightarrow i, l \leftrightarrow j \}$.
Using the above mode expansion, it is straightforward to compute the graviton
4-point functions, leaving them in terms of 6D virtual momenta integrals. 
After summing up two Wick contractions in each correlation function, we find\footnote{Henceforth, we omit the time dependence of the Hubble parameter which should be understood to be that at horizon crossing.} 
\bea
\label{g1}
\langle 0 | h_{kl} (x_1) h_{ij} (x_2) h_{mn} (\vec{x}, 0) h_{mn} (\vec{x}', 0 ) | 0 \rangle &=& 
\frac{8}{(2\pi )^6} \frac{H^4}{M^4_p} \int d^3 s d^3 s' \,\,\,
e^{i \vec{s} \cdot (\vec{x}_1 - \vec{x} )} e^{i \vec{s}' \cdot (\vec{x}_2 - \vec{x}' )}
e^{- i s \tau_1 - i s' \tau_2 }  \cr
&&  \times \frac{(1+i s \tau_1 ) (1+ i s' \tau_2 )}{s^3 s'^3}   \sum_{\lambda, \lambda'} \epsilon_{kl} (\hat{s}, \lambda ) \epsilon^*_{mn} (\hat{s}, \lambda ) \epsilon_{ij} (\hat{s}', \lambda' ) \epsilon^*_{mn} (\hat{s}', \lambda' ),  \nonumber  \\ \\ 
\label{g2}
\langle 0 | h_{kl} (x_1)  h_{mn} (\vec{x}, 0) h_{mn} (\vec{x}', 0 ) h_{ij} (x_2) | 0 \rangle &=& 
\frac{8}{(2\pi )^6} \frac{H^4}{M^4_p} \int d^3 s d^3 s' \,\,\,
e^{i \vec{s} \cdot (\vec{x}_1 - \vec{x} )} e^{i \vec{s}' \cdot (\vec{x}_2 - \vec{x}' )}
e^{- i s \tau_1 + i s' \tau_2 }  \cr
\label{gravitonM}
&&  \times \frac{(1+i s \tau_1 ) (1- i s' \tau_2 )}{s^3 s'^3}  \sum_{\lambda, \lambda'}  \epsilon_{kl} (\hat{s}, \lambda ) \epsilon^*_{mn} (\hat{s}, \lambda ) \epsilon_{ij} (\hat{s}', \lambda' ) \epsilon^*_{mn} (\hat{s}', \lambda' ).  \nonumber \\
\eea

\subsection{A schematic outline of the one-loop computation}
\label{sec:Order}

In the following, we present the schematic outline of our one-loop computation in particular the order of integrations
in $\int d^3x_1 d^3x_2 \,\, \langle 0 | H_{int} (x_1) H_{int} (x_2) h_{mn} (\vec{x}, 0 ) h_{mn} (\vec{x}', 0 ) | 0 \rangle$
and \newline
$\int d^3x_1 d^3x_2 \,\, \langle 0 | H_{int} (x_1) h_{mn} (\vec{x}, 0 ) h_{mn} (\vec{x}', 0 )  H_{int} (x_2) | 0 \rangle$. 
The full one-loop result is eventually obtained after performing the time integral contours following \eqref{SK1} and performing
appropriate regularizations of the momenta integrals.

\begin{enumerate}[(i)]
\item
Each matter field 4-point function can be expressed in terms of 
a six-dimensional virtual momenta integral in the form 
$$
\langle \mathcal{G}^{kl} (x_1) \mathcal{G}^{ij} (x_2 ) \rangle = 
\int d^3p_1 d^3p_2\, e^{
i( \vec{p}_1 + \vec{p}_2 ) \cdot (\vec{x}_1 - \vec{x}_2 )
} G^{klij} (\vec{p}_1, \vec{p}_2, \tau_1, \tau_2 ),
$$
where $G^{klij}$ is a function of momenta and time that depends on the specific Hamiltonian 
and the mode wavefunctions of the matter field. It is a sum of two Wick contractions. 

Taking into account the
phase factor $e^{i \vec{s} \cdot (\vec{x}_1 - \vec{x} )} e^{i \vec{s}' \cdot (\vec{x}_2 - \vec{x}' )}$
in the graviton 4-point functions in \eqref{gravitonM}, 
we first integrate over $\vec{x}_{1,2}$ to obtain the delta functions
$$
(2\pi )^6  \delta^3 ( \vec{s} + \vec{p}_1 + \vec{p}_2 ) \delta^3 (\vec{s}' - \vec{p}_1 - \vec{p}_2 )
$$
together with a phase factor $e^{i (\vec{p}_1 + \vec{p}_2 ) \cdot (\vec{x} - \vec{x}' )}$.

\item Then we integrate over $\vec{s}, \vec{s}'$ in the graviton 4-point function using the above delta functions. 
The integrand is now simply a function of virtual momenta $\vec{p}_1, \vec{p}_2$.

\item Finally, we perform a Fourier transform $\int d^3x e^{i\vec{q} \cdot (\vec{x} - \vec{x}' )} 
e^{i (\vec{p}_1 + \vec{p}_2 ) \cdot (\vec{x} - \vec{x}' )}
 = (2\pi )^3 \delta^3 (\vec{q} + \vec{p}_1 + \vec{p}_2 )$,
where $\vec{q}$ is the external momentum.

This leads to the eventual integral measure
 $\int d^3 p_1 \int d^3 p_2 \delta^3 (\vec{q} + \vec{p}_1 + \vec{p}_2 )$. 
For computational convenience, we can write this as\footnote{This is valid as long as the integrand
is a function of the magnitudes of $\vec{p}_1, \vec{p}_2,\vec{q}$. Using the delta function, we can first integrate
over $p_2$. Since the $p$-integral is over the entire momentum space, we can choose $\vec{q}$
to lie along the $z$-axis. Together with the relation $p_2^2 = p^2_1 + q^2 +2p_1 q \cos (\theta)$ 
which implies that $p_2 d p_2 = p_1 q d(\cos (\theta ))$, we are led to \eqref{integralform} after integral over the
azimuthal angle in $p_1$-momentum space. We note that the upper and lower limits of the $p_2$-integral correspond to $\theta =0,\pi$ respectively.}
\be
\label{integralform}
\int d^3 p_1 \int d^3 p_2 \delta^3 (\vec{q} + \vec{p}_1 + \vec{p}_2 )
= \frac{2\pi}{q} \int^\infty_0 dp_1 \int^{p_1+q}_{|p_1-q|} dp_2\,\, p_1 p_2
\ee
We will find that these integral diverge yet a suitable regularization 
parametrized by $H$ - the Hubble parameter at horizon exit and $\mu$ -  the renormalization constant. 
This enables us to extract the finite logarithmic term in the one-loop correction which can be eventually expressed in the form
\be
I = \frac{1}{(2\pi)^2} \left( \frac{H}{M_p} \right)^4 \frac{1}{q^3} F\left( \frac{H}{\mu} \right)
\ee
for some function $F\left( \frac{H}{\mu} \right)$. 
\end{enumerate}
\noindent
In the following section, we will perform the above computations for various matter fields.
Since the procedure is identical for all, we introduce a few other symbols to 
organize our presentation of results. 
Denoting the one-loop correction term by $I_L$, 
we distinguish between the two time integral contours as follows.
\bea
I_L &=& \frac{2\pi}{q}  \int^\infty_0 dp_1 \int^{p_1+q}_{|p_1-q|} dp_2\,\, p_1p_2 \,\, 
\mathcal{G} (p_1,p_2,q) 
\left[ -2 \text{Re} (F_1 (p_1, p_2, q) ) + F_2 (p_1, p_2, q) \right]\cr
\label{iloop}
& \equiv &
\frac{1}{(2\pi )^2} \left( \frac{H}{M_p} \right)^4 \frac{1}{q^3} \left(
I_1 + I_2
\right)
\eea
where $F_1 (p_1, p_2, q), F_2 (p_1, p_2, q)$ are the time integrals involving the time-dependent functions
that appear in the fields' mode wavefunctions, and 
$\mathcal{G} (p_1, p_2, q)$ capture all other functions and constants obtained after contracting the spacetime indices in the product
$$G^{klij} (\vec{p}_1, \vec{p}_2, \tau_1, \tau_2 )
 \left( 
\sum_{\lambda, \lambda'} \epsilon_{kl} (\hat{q}, \lambda) \epsilon^*_{mn} (\hat{q}, \lambda ) \epsilon^*_{ij} 
(\hat{q}, \lambda' ) \epsilon_{mn} (\hat{q}, \lambda' ) 
\right).$$
Some details of the computation will be presented for each matter field. The momenta integrals yield
divergent results and must be regularized, after which 
$$
I_1 + I_2 \rightarrow F(H/\mu ).
$$
Specifically, both cutoff and dimensional regularization leads to $F(H/\mu )$ being some constant multiplied to $\log (\frac{H}{\mu} )$.

\subsection{Regularization methods}
\label{sec:Regular}

In the following, we explain the methods used to regularize the divergent momenta integrals in \eqref{iloop}. 
For all matter fields we consider, we employ both dimensional regularization and cutoff regularization and in all
cases, we obtain identical finite one-loop correction terms.

\subsubsection{Dimensional regularization}

For dimensional regularization, 
we begin by 
writing the spatial dimensionality as
$
d = 3+ \delta,
$
and noting that the angular integration should generalize as 
\be
\label{angular}
\int d\Omega_{2+\delta} = \int^\pi_0 d\Theta\, \, \sin^{1+\delta} \Theta \times \text{Vol} (S_{1+\delta} ) = 
\int^\pi_0 d\Theta \sin^{1+\delta} \Theta \times  \frac{(2\pi)\pi^{\delta/2}}{\Gamma \left(1 + \frac{\delta}{2}\right)}. 
\ee
The integration measure then reads 
\be
\label{comDim}
 \int d^{3+\delta}p_1 \int d^{3+\delta} p_2 \,\, \delta^{3+\delta} ( p_1 + p_2 + q)
=2\pi
q^{3+\delta}  \left(   \frac{\pi^{\delta /2}}{\Gamma (1 + \delta /2)}    \right)
\int^{\infty}_0 dp_1\,\, p^{\delta}_1 \, \int^{p_1+1}_{|p_1-1|} dp_2\,\, \sin^\delta \Theta 
\ee
where we have expressed the integrand variables in units of $q$ and  
$$
\sin^\delta \Theta = \left[  1 - \left( \frac{(p_2)^2 - p^2_1 -1}{2p_1}     \right)^2   \right]^{\delta/2}.
$$
There is also the spacetime integral measure $\int d^4x_{1,2} a^4 (\tau_{1,2} )$ which is lifted to 
be $\int d^{4+\delta} x_{1,2} a^{4+\delta} (\tau_{1,2} )$. Up to first-order in $\delta$, 
\be
\label{scaleDim}
a^{\delta} (\tau ) = 1 - \delta \text{Log} (-H \tau ) + \mathcal{O}(\delta^2 ),
\ee
and we note that we have a product of two such measures in the correlation function that involves
a product of two interaction Hamiltonians. The integrand of \eqref{iloop} involves mode wavefunctions
and those running in the loop can be analytically continued to the corresponding solutions in higher dimensions. 
In this paper, we consider minimally coupled scalars, Dirac fermions, and the abelian gauge field. 
Below, we present their higher-dimensional modes which are crucial for dimensional regularization.

For the scalar field, the analytic continuation of the scalar mode wavefunctions
\be
\label{scalarHigher}
\chi_k (\tau ) \sim \frac{H^{1+ \delta /2}  (-k \tau )^{(3+\delta )/2}}{k^{(3+\delta )/2}} H^{(1)}_{(3+\delta )/2} (-k\tau )
\ee
where $H^{(1)}_{(3+\delta)/2} (-k \tau )$ is the Hankel function of the first kind. For the minimally coupled
scalars, the terms in $H_{int}$ which give non-vanishing contribution to the one-loop correction 
are quadratic in the fields, so expanding in the parameter $\delta$ yields 
$$
4 \times \frac{1}{2} \delta \log (-H \tau ).
$$
The form of \eqref{scalarHigher} is, up to some constants, identical for the graviton mode wavefunctions $h_k (\tau )$
in \eqref{gravitonmodes}. Since there is one in each of $H_{int}$, together with the scalar mode wavefunctions, we 
have altogether a factor of 
\be
6 \times \frac{1}{2} \delta \log (-H \tau ).
\ee
This turns out to be identical for the other matter fields that we consider in this paper, each of them having a
interaction Hamiltonian that is linear in $h_{kl}$ and quadratic in the fields (seagull vertices as depicted in Figure~\ref{fig:f2}
don't contribute and are treated in Section~\ref{sec:Seagull}). 

For the fermions, in $d+1$-dimensional de Sitter, the Dirac equation turns out to read 
$$i \left(\gamma^\mu \partial_\mu - \frac{d}{2\tau} \gamma^0 \right)\Psi = 0,$$ which admits the $d-$dimensional spinor wavefunction 
\be
\Psi (\vec{x}, t) = (-H\tau )^{\frac{d}{2}}  \int d^dk\,\, \sum_s \, e^{i\vec{k} \cdot \vec{x}} 
\left[   U_{\vec{k}, s} (t) a_{\vec{k},s} + V_{-\vec{k}, s} (t) \beta^\dagger_{-\vec{k}, s}
\right], 
\ee
where $U_{\vec{k},s }, V_{\vec{k},s}$ are spinors with definite conformal momenta in $d+1$-dimensional Minkowski spacetime. Expanding 
$d=3+\delta$, one obtains a correction factor of $\frac{1}{2} \delta \log (-H\tau) $, similar to what arises
from the analytic continuation of scalar mode wavefunctions. 

For the gauge fields, we checked that the free
$d-$dimensional Maxwell equations in conformal coordinates imply that the modes $A_k (\tau )$ satisfy 
\be
\frac{d^2A_k}{d\tau^2} + k^2 A_k - (d-3)\frac{1}{\tau} \frac{dA_k}{d\tau} = 0.
\ee
In the case of $d=3$, we obtain plane waves which can be normalized with Bunch-Davies condition. 
For generic $d= 3+\delta$,
we find the solution  
\be
A_k \sim (-H\tau )^{\frac{1+\delta}{2}} H^{(1)}_{\frac{1+\delta}{2}} (-k\tau),
\ee
where $H^{(1)}_v (-k\tau )$ denotes the Hankel function of the first kind. 
Once again expanding $d = 3+\delta$, one obtains the same correction factor of $\frac{1}{2} \delta \log (-H\tau) $.

To summarize, for all matter fields, the analytic continuation of the mode wavefunctions as well as the scale factor
in each integration of $H_{int}$ over all spacetime implies the following term that is first-order in the regularization parameter $\delta$:
\be
\label{integrandHigher}
\left( \frac{4+2}{2} - 2 \right) 
\delta \log (-H\tau)
\ee
Although we shall see later that seagull vertices do not contribute to the one-loop correction, we note in passing that
a similar calculation gives 
$
\left( \frac{2+2}{2} - 1 \right) 
\delta \log (-H\tau)
= 
\delta \log (-H\tau)
$
which is identical to \eqref{integrandHigher}. Finally, taking into account the $q^\delta$ term
in \eqref{comDim} (the other $\delta$-dependent terms in \eqref{comDim} will only
yield unimportant constants), we can derive the expression for the finite one-loop logarithmic correction 
in dimensional regularization which is the finite part of 
\be
\label{resi}
I_L \sim \frac{1}{(2\pi )^2} \left( \frac{H}{M_p} \right)^4 \frac{1}{q^3} 
 \left( 1 + \delta \log \left( \frac{q}{\mu} \right) \right) \left(- \frac{J_1 + J_2}{\delta} + \ldots \right)
\left( 1 + \delta \log \left( \frac{H}{q} \right)  \right) ,
\ee
where we have also invoked the fact that the time-integrals are dominated by the time of horizon exit,
and $J_{1,2}$ denote the residues of the momenta integral in \eqref{iloop} 
in each of the time integral contour after scaling all virtual momenta in units
of $q$.  
To pick up this residue, we find it convenient to perform a coordinate transformation as follows. 
\be
\label{coordinateT}
P = p^\delta_1.
\ee
In \eqref{iloop}, we first integrate over $p_2$ without encountering any divergences. Then
we perform the coordinate transformation in \eqref{coordinateT}. The integral from $p_1=0$ to $p_1=q$
does not contribute to the logarithmic correction so we focus on the remaining domain of integration. 
Scaling all virtual momenta in units of $q$, the one-loop correction is the finite part of the following expression
\bea
I_L &\sim& \frac{1}{(2\pi )^2} \left( \frac{H}{M_p} \right)^4 \frac{1}{q^3} \int^\infty_1 \, dP\, \frac{1}{\delta}
\left[ 
J_1 (P, \delta) + J_2 (P, \delta ) 
\right]
\left( 1 + \delta \log \left(  \frac{H}{\mu} \right) \right) \cr
\label{DimRegular}
&=&- \frac{1}{(2\pi )^2} \left( \frac{H}{M_p} \right)^4 \frac{1}{q^3} \left[ J_1 + J_2  \right] \log 
\left(  \frac{H}{\mu} \right),
\eea
where $J_1 (P, \delta), J_2 (P, \delta)$ denote the two functions arising from each time integral contour
in \eqref{iloop} and $J_{1,2}$ in \eqref{resi} are then simply the constant terms 
that can be read off from the Taylor expansion of $J_1 (P, \delta) + J_2 (P, \delta)$
in the small variable $P^{-\frac{1}{\delta}}$.  For all matter fields we consider in this paper,
the one-loop correction is of this form. Finally, we note that the other $\delta$-dependent terms in \eqref{comDim}, apart from $q^\delta$,  do not contribute to this logarithmic running apart from unimportant numerical constants. 
Explicitly, the factor 
$$
\frac{\pi^{\delta /2}}{\Gamma ( 1 + \frac{\delta}{2} ) } \approx 1 +\delta \left( \frac{\gamma}{2} + \frac{1}{2} \text{Log} ( \pi ) \right) + \ldots,
$$
whereas expanding the $\sin^\delta \Theta$ term introduces a term 
$
\text{Log}  \left( 1 - \frac{(p^2_2-p^2_1-1)^2}{4p^2} \right)
$
in the integrand, which gives rise to an unimportant numerical constant. 
In the subsequent sections, we present some details of $J_1 (P, \delta), J_2 (P, \delta)$ for each matter field
and thus also the one-loop logarithmic correction.


\subsubsection{A covariant cutoff regularization}

Another regularization method that we adopt to derive our result is that of cutoff regularization as
first discussed in \cite{Weinberg} and later refined in \cite{Senatore}. The method has been applied to a number of different contexts in \cite{Senatore} and we will later see that it yields the same results as those obtained by
dimensional regularization, thus furnishing a good consistency check.

We first begin from \eqref{iloop} which we reproduce here for convenience.
\bea
\label{iloopref}
I_L &=& \frac{2\pi}{q}  \int^\infty_0 dp_1 \int^{p_1+q}_{|p_1-q|} dp_2\,\, p_1p_2 \,\, 
\mathcal{G} (p_1,p_2,q) 
\left[ -2 \text{Re} (F_1 (p_1, p_2, q) ) + F_2 (p_1, p_2, q) \right] \cr
& \equiv&  \frac{1}{(2\pi)^2} \left( \frac{H}{M_p}  \right)^4 
\frac{1}{q^3} (I_1 + I_2). 
\eea
The functions $F_1, F_2$ are derived by completing the time-integrals. To regularize $I_L$ following
\cite{Senatore}, we first note that the momentum integral should be regulated with a covariant
cutoff as follows
\be
\int d^3 p \rightarrow \int^{\Lambda a (\tau)} d^3 p,
\ee
where $\Lambda$ is the fixed physical cutoff such that the physical momentum $p_{phy} = \frac{p}{a(\tau)} < \Lambda$. 

As explained in \cite{Senatore}, the time integral needs to be regulated too,
by taking $\tau \rightarrow \tau (1 + \frac{H}{\Lambda} )$. \footnote{In terms of the comoving time $t$, after taking the limit of large $\Lambda$
and in the slow-roll approximation where we take $a(\tau) \approx -1/(H\tau)$,
this translates to the cutoff $t \rightarrow t - \frac{1}{\Lambda}$.} Since one limit of the momentum integral 
is time-dependent, it is natural to let it precede temporal ones, and thus the regularized version of 
\eqref{iloopref} reads 
\bea
\label{tildeI1}
\tilde{I}_1  &=& 
\int^0_{-\infty_+} d\tau_2 \int^{\tau_2 (1 + \frac{H}{\Lambda} ) }_{-\infty_+} d \tau_1 
\int^{\Lambda a(\tau_1)}_0 d p_1  
 \int^{p_1+q}_{|p_1-q|} dp_2 \, \,\,\, \tilde{f}_1 (p_1,p_2,\tau_1,\tau_2), \\
\label{tildeI2}
\tilde{I}_2 &=& 
\int^0_{-\infty_-} d\tau_2 \int^{0}_{-\infty_+} d \tau_1 
\int^{\Lambda a(\tau_1)}_0 d p_1 
\,\, \int^{p_1+q}_{|p_1-q|} dp_2 \,\,\,\,  \tilde{f}_2 (p_1,p_2,\tau_1,\tau_2),
\eea
where we have succintly let $\tilde{f}_{1,2}$ denote the full integrands and explicitly restored the time-integrals 
as first introduced in \eqref{SK1}. 

In the next Section, we will evaluate the one-loop correction for each matter fields using dimensional
regularization as well as the covariant cutoff method. With regards 
to the latter, we find that it seems technically difficult to extract the logarithmic term by 
bluntly alluding to the integration order specified in \eqref{tildeI1} and \eqref{tildeI2}, mainly due
to the fact that it turns out we can't perform certain integrals analytically in our case. 
To circumvent this problem, just as in any suitable
multi-dimensional integral, we attempt to see if the problem becomes tractable once we rewrite
these equations by re-ordering the integrals equivalently as follows. Let's begin with 
terms in $\tilde{I}_1$. 
First we shift the $p_1$ integral past the $\tau_1$ integral on its left by writing
\be
\label{firstswitch}
\int^{\tau_2 (\frac{H}{\Lambda} + 1)}_{-\infty_+} d\tau_1 \,\,
\int^{-\frac{\Lambda}{H\tau_1}}_0 dp_1 
= \int^{-\frac{\Lambda}{H\tau_2}}_0 dp_1 
\int^{\tau_2 (\frac{H}{\Lambda} + 1)}_{-\left(  \frac{\Lambda}{Hp_1} \right)_+} d\tau_1,
\ee
where we have retained the infinitesimal analytic continuation associated with the $i\epsilon$ prescription.
Next, we shift the $p_1$ integral past the second time integral as follows. 
\be
\label{secondswitch}
\int^{0}_{-\infty_+} d\tau_2
\int^{-\frac{\Lambda}{H\tau_2}}_0 dp_1  = 
\int^{\infty}_0 dp_1 \,\, \int^{0}_{-\left(  \frac{\Lambda}{Hp_1} \right)_+} d\tau_2.
\ee
The other momentum
integral doesn't involve time in their limits so it can be shifted past the temporal integrals without modification if we wish. 
For notational clarity,  we will place it on the right of the $p_1$-integral from now. Thus, \eqref{tildeI1} now reads
\be
\label{newcutoff1}
\tilde{I}_1 = 
\int^{\infty}_0 dp_1 \,\, \int^{p_1+q}_{|p_1-q|} dp_2 \,\,
\int^{0}_{-\left(  \frac{\Lambda}{Hp_1} \right)_+} d\tau_2 \int^{\tau_2 (\frac{H}{\Lambda} + 1)}_{-\left(  \frac{\Lambda}{Hp_1} \right)_+} d\tau_1
\,\,  \tilde{f}_1 (p_1,p_2,\tau_1,\tau_2).
\ee
Similarly, for $\tilde{I}_2$ in \eqref{tildeI2}, we have 
\be
\label{newcutoff2}
\tilde{I}_2 =
\int^{\infty}_0 dp_1 \,\, \int^{p_1+q}_{|p_1-q|} dp_2 \,\,
\int^{0}_{-\left(  \frac{\Lambda}{Hp_1} \right)_-} d\tau_2 \int^{0}_{-\left(  \frac{\Lambda}{Hp_1} \right)_+} d\tau_1
\,\,\,\tilde{f}_2 (p_1,p_2,\tau_1,\tau_2).
\ee
In the limit of infinite $\Lambda$ cutoff, both \eqref{newcutoff1} and \eqref{newcutoff2} reduce to the unregulated expression in \eqref{iloopref}
of which time-integrals are defined in \eqref{SK1}. We note that the presence of momentum-dependent limits in the time integrals arise fundamentally from the time-dependent limit of the momentum integral in \eqref{tildeI1} and \eqref{tildeI2}. In principle, there is no a priori advantage in picking the form in \eqref{newcutoff1} and \eqref{newcutoff2} over the former expressions. For example, in \cite{Senatore}, a number of results were obtained via \eqref{tildeI1} and \eqref{tildeI2}. 

There is still one remaining subtlety left in evaluating the one-loop logarithmic correction term 
from \eqref{newcutoff1} and \eqref{newcutoff2}. A direct computation of them yields a divergent quantity, even for a large finite $\Lambda$ (a feature which expectedly characterizes the equivalent integrals in the order expressed in \eqref{tildeI1} and \eqref{tildeI2} too). To proceed, we first decompose the $p_1$-integral into two parts as follows:
\be
\label{mnt}
\int^\infty_0 dp_1 = \int^{\frac{\Lambda q}{H}}_0 dp_1 +\int^\infty_{\frac{\Lambda q}{H}} dp_1,
\ee
where the second piece contains the UV divergence that characterizes the Minkowski limit as well. We note that in \eqref{secondswitch}, if we restrict the time-domain to be bounded by the time of horizon-crossing
($q = aH \approx -1/\tau_2$), then the momentum integral is bounded by $\frac{\Lambda q}{H}$ which is the separation value picked in \eqref{mnt}. We find that the time integrals containing exponential terms\footnote{They are presented explicitly in \eqref{fermionTime}, \eqref{scalarTime} and \eqref{gaugeTime} in Section \ref{sec:Matter} .} are dominated by behavior surrounding the period of horizon-crossing --- an observation which was made in similar contexts in \cite{Weinberg,Senatore}.\footnote{In \cite{Weinberg}, it was argued that the notion of horizon-crossing in the internal space is only tenable provided the temporal integrals
are dominated by behavior surrounding $p\sim q \sim aH$.} While the second piece of \eqref{mnt} harbors the UV divergence, the first piece contains the one-loop correction term we are seeking. From this point onwards, we will focus our computation on this part of the momentum integral. 

At early times, the convergence of the time integrals is characterized by $\Lambda$. 
As emphasized earlier, the momentum (and $\Lambda$-dependence) of the limits in \eqref{tildeI1} and \eqref{tildeI2} arises from the fundamental time dependence of the momentum cutoff and switching the order of integration. It's thus important to study if they could potentially contribute to the one-loop logarithmic term. In Section \ref{sec:Matter} where we present the details of the computation, we present explicitly the leading order terms arising from these limits and find that they ultimately do not contribute (see discussion surrounding \eqref{fermionTime} and \eqref{scalarTime}). Generally, it turns out that they are of the form $w(p_1,p_2) e^{-v(p_1,p_2)\epsilon \frac{\Lambda}{H}} \left(
\frac{\Lambda}{H}\right)^n$ for some positive integer $n$, where we have restored the analytic continuation parameter $\epsilon$.

\section{One-loop logarithmic corrections due to massless spectator fields}
\label{sec:Matter}

In this section, we evaluate the functions $\mathcal{G} (p_1, p_2, q), \mathcal{F} _1 (p_1, p_2, q), \mathcal{F} _2 (p_1, p_2, q)$ in \eqref{iloop} with both regularization methods which, as we shall see, will yield identical results for the 
finite logarithmic term. For dimensional regularization, we compute $J_1, J_2$ in \eqref{DimRegular} following
the discussion surrounding \eqref{coordinateT} whereas for cutoff regularization, we evaluate the logarithmic
term using \eqref{newcutoff1} and \eqref{newcutoff2}. The scheme of computation follows the method explained in the preceding section \eqref{sec:Order}.

\subsection{Dirac Fermions}
\label{sec:Fermion}

We consider massless Dirac fermions coupled to the inflationary background as described by the action 
\be
S_{fermion} = \int d^4x \sqrt{-g} \frac{i}{2} 
\left(
\overline{\varphi} \Gamma^\mu \mathcal{D}_\mu \varphi -  \mathcal{D}_\mu \overline{\varphi} \Gamma^\mu \varphi 
\right)
\ee
where $\Gamma^\mu = e^\mu_a \gamma^a$ are the Dirac matrices on the FRW spacetime with vierbein 
$e^\mu_a = \frac{1}{a(\tau)} \delta^\mu_a$, $\gamma^a$ are Dirac matrices on Minkowski spacetime,
and $\mathcal{D}_\mu = \partial_\mu + \Omega_\mu$ is the covariant derivative with spin connection $\Omega$. 
For the FRW background, $\mathcal{D}_0 = \partial_0, \mathcal{D}_k = \partial_k + \frac{H}{2} \gamma_k \gamma^0$. 

With the metric perturbation switched on, one can expand the Lagrangian to first-order in metric fluctuation $h_{ij}$ and
after a Legendre transform, we obtain the interaction Hamiltonian
\be
H= -\frac{i}{2} \int d^3x \, a^3 (\tau )\, h_{ij} 
\left[  
\overline{\varphi} \gamma^i \partial_j \varphi - \partial_i \overline{\varphi} \gamma^j \varphi
\right].
\ee
The spinor fields can be expanded in terms of modes labelled by spin index $\lambda$
\be
\varphi = \int d^3p e^{i \vec{p} \cdot \vec{x} } \sum_\lambda  \left(
a_{\vec{p}, s} U_{\vec{p}, s} (\tau ) + b^\dagger_{-\vec{p}, s} V_{-\vec{p}, s} (\tau )
\right),
\ee
with the oscillator algebras being
 $$\{ a_{\vec{p}, \lambda}, a^\dagger_{\vec{p}', \lambda' } \} = \delta^3 (\vec{p} - \vec{p}') \delta_{\lambda \lambda'},\,\,\, \{ b_{\vec{p}, \lambda}, b^\dagger_{\vec{p}', \lambda' } \} = \delta^3 (\vec{p} - \vec{p}') \delta_{\lambda \lambda'},
$$ 
and the normalization condition 
\be
U_{\vec{p}, s} \overline{U}_{\vec{p}, s} = 
V_{\vec{p}, s} \overline{V}_{\vec{p}, s} = \frac{\gamma^\mu p_\mu}{2 (2\pi )^3 a^3 (\tau ) p}.
\ee
The spinor field 4-point function can be expressed as
\be
\langle
\overline{\varphi} (x_1) \gamma^l \partial_k \varphi (x_1) 
\overline{\varphi} (x_2) \gamma^j \partial_i \varphi (x_2) 
\rangle
=
 \int
d^3p d^3p'\, e^{i(\vec{p}+\vec{p}')(\vec{x}_1 - \vec{x}_2)} p'^k p^i  e^{-i (p+p') (\tau_1 - \tau_2)} \sum_{s,s'} 
 \overline{U}_{\vec{p},s} \gamma^j V_{\vec{p}', s'} \overline{V}_{\vec{p}', s'} \gamma^l U_{\vec{p},s}.
\ee
Simplifying further using
\be
 \sum_{s,s'} 
 \overline{U}_{\vec{p},s} \gamma^j V_{\vec{p}', s'} \overline{V}_{\vec{p}', s'} \gamma^l U_{\vec{p},s} = 
\sum_{s,s'} \text{Tr} \left( \gamma^l \left(  \frac{\slashed{p}}{2(2\pi)^3p}   \right) \gamma^j \left(  \frac{\slashed{p'}}{2(2\pi)^3p'}  \right) \right)
=\frac{1}{4(2\pi)^6 p p'} \text{Tr} \left(  
\gamma^l \slashed{p} \gamma^j \slashed{p}' 
\right),
\ee
we can contract it with the graviton's polarization tensors to obtain
\bea
\mathcal{G} (p_1, p_2, q) &=& 
- N (p^k_2 - p^k_1)( p^i_1 - p^i_2) \text{Tr} (\gamma^l \slashed{p_1} \gamma^j \slashed{p}_2 ) 
 \sum_{\lambda, \lambda'}
\epsilon_{kl} (q, \lambda) \epsilon^*_{mn} (q, \lambda ) \epsilon^*_{ij} (q, \lambda' ) \epsilon_{mn} (q, \lambda' ) \cr
&=& 
N (2p^k_1+q^k)( 2p^i_1 +q^i) \left[
(p_1 p_2 - \vec{p}_1 \cdot \vec{p}_2 ) 4\delta^{lj} +  4( p^l_1 p^j_2 +p^j_1 p^l_2 ) \right] \cr
&&\qquad  \times
2 \left( \delta_{ki} \delta_{lj} - \delta_{ki} \hat{q}_l \hat{q}_j - \delta_{lj} \hat{q}_k \hat{q}_i + \hat{q}_k
\hat{q}_l \hat{q}_i \hat{q}_j \right)\nonumber \cr
&=& -N \frac{4}{q^4} 
\left(p^2_1 + 2p_1 p_2 + p^2_2 - q^2)^2 (p^4_1 - 4p^3_1 p_2 + 6p^2_1 p^2_2 -  4p_1 p^3_2 + p^4_2 - q^4 \right)
\eea 
where $N= \frac{4}{(4\pi)^3 q^6} \frac{H^4}{M^4_p}$ and we have invoked the useful identity 
$
\text{Tr} (\gamma^a \gamma^b \gamma^c \gamma^d ) = 4 \left(  
\eta^{ab} \eta^{cd} - \eta^{ac} \eta^{bd} + \eta^{ad} \eta^{bc} 
\right)
$
to write
\bea
\text{Tr} \left( \gamma^l \gamma^\mu \gamma^j \gamma^\beta \right) p_\mu p'_\beta &=& 
p p' \text{Tr} \left(  \gamma^l \gamma^0 \gamma^j \gamma^0 \right) + p^b p'^d 4\left( \eta^{lb} \eta^{jd} - \eta^{lj} \eta^{bd} + \eta^{ld} \eta^{bj} 
\right) \cr
&=& (pp' - \vec{p} \cdot \vec{p}' ) 4\delta^{lj} +  4( p^l p'^j +p^j p'^l ).
\eea
We absorb all time-dependent functions into $F_1(p_1, p_2, q), F_2(p_1, p_2, q)$ which represent the two time integration contours. In this case,
it reads 
\bea
F_1(p_1, p_2, q) &=& \int^{0}_{-\left(  \frac{\Lambda}{Hp_1} \right)_+} d\tau_2 \int^{\tau_2 (\frac{H}{\Lambda} + 1)}_{-\left(  \frac{\Lambda}{Hp_1} \right)_+} d\tau_1\,\,\,
 e^{-i(p_1 + p_2)(\tau_1 - \tau_2)} 
(1+iq \tau_1 )(1+i q \tau_2 ) e^{-iq (\tau_1 + \tau_2)} \cr
\label{fermionTime}
F_2(p_1, p_2, q) &=& \int^{0}_{-\left(  \frac{\Lambda}{Hp_1} \right)_-} d\tau_2 \int^{0}_{-\left(  \frac{\Lambda}{Hp_1} \right)_+} d\tau_1\,\,\,e^{-i(p_1 +p_2)(\tau_1 - \tau_2)} 
(1+iq \tau_1 )(1-i q \tau_2 ) e^{-iq (\tau_1 - \tau_2)} \nonumber \\
\eea
In the early-time regime, the convergence of the time integrals is parametrized by $\Lambda$ as follows. In the large $\Lambda$ expansion, 
the leading-order term goes as 
$$
F_1 \sim \left(   
\frac{q(p_1+p_2-2q)}{p_1(q+p_1+p_2)(p_1+p_2-q)^2}
\right)
e^{-\epsilon \Lambda \frac{q+p_1 +p_2}{Hp_1}} \frac{\Lambda}{H} + \ldots +
\left(   
\frac{q^2}{p^2_1 q (q+p_1+p_2)}
\right)
e^{-2\epsilon \Lambda \frac{q}{Hp_1}} \frac{\Lambda^2}{H^2} + \ldots,
$$
where we have displayed the $\Lambda$-dependent leading-order terms corresponding to the lower and
upper limits of the $\tau_1$-integral in $F_1$ respectively. Similarly for $F_2$, the $\Lambda$-dependent leading-order term goes as
$$
F_2 \sim \left(
\frac{q(2q+p_1+p_2)}{p_1(q+p_1+p_2)^3}
\right)
e^{-\epsilon \Lambda  \frac{(q+p_1 + p_2)}{Hp_1}} \frac{\Lambda}{H} + \ldots
$$
These expressions show that the $\Lambda$-dependent integral limits of \eqref{fermionTime} do not enter
into the final expression of the logarithmic term we are seeking.

After performing the time integrals, one can proceed to evaluating \eqref{iloop}. Directly performing
the momenta integration with a cutoff yields
($\tilde{L} = L/q$ )
\be
\label{oneLoopF}
I_1+I_2 = \frac{32}{5} \ \text{Log} (\tilde{L} )- \frac{32}{5} \left[ -1 + \tilde{L} \left( -5 + 2\tilde{L} (-5-5\tilde{L} + 2\tilde{L}^3 ) \right)  \right] \text{Log} (1 + \tilde{L}^{-1} ) + \ldots 
\sim -\frac{32}{5} \ \text{Log} (\frac{q}{\mu} )
\ee
where the ellipses refer to terms which are polynomial in $\tilde{L}$ and those which vanish with infinite $\tilde{L}$.
 The logarithmic term yields the finite one-loop correction 
which is thus 
\be
\label{Fresult}
I_L = -\frac{H^4 (\tau_q ) }{(2\pi)^2 M^4_p q^3} \times \frac{32}{5} \text{Log} \left( \frac{H}{\mu} \right).
\ee
We note that the first time integral contour does not contribute. This is mirrored in the 
alternative method of dimensional regularization. Following \eqref{DimRegular},
 we have from each time integration contour
\be
\frac{1}{\delta}  \int^\infty_1 dP \, J_1 (P, \delta ) 
 = \frac{4}{\delta}  \int^\infty_1 dP \, 
\frac{16}{P^{-\frac{4}{\delta} }} - \frac{16}{5 P^{-\frac{3}{\delta} }} - \frac{208}{35 P^{-\frac{2}{\delta} }} + \frac{316}{105 P^{-\frac{1}{\delta} }} + \ldots
\ee
and
\be
\frac{1}{\delta}  \int^\infty_1 dP \, J_2(P, \delta )  
=  \frac{4}{\delta} \int^\infty_1 dP \, 
\frac{16}{5 P^{-\frac{3}{\delta} }} - \frac{316}{105 P^{-\frac{1}{\delta} }} + \frac{8}{5} - \frac{32 P^{-\frac{2}{\delta} }}{105}
+ \ldots
\ee
Combining both $I_1$ and $I_2$, we obtain the identical one-loop logarithmic correction \eqref{Fresult}
computed in cutoff
regularization.


\subsection{Minimally coupled scalars}
\label{sec:Scalar}

We consider the minimally coupled scalar with action
\be
S_{scalar} = \frac{1}{2} \int d^4x  \sqrt{-g} g^{\mu \nu} \partial_\mu \phi \partial_\nu \phi .
\ee
Switching on the metric perturbation and keeping to first-order, after performing a Legendre transform, we obtain the interaction 
Hamiltonian to read
\be
H_{int} = \frac{1}{2} \int d^3x\, a^2 (\tau )\,
 h^{ij} \partial_i \phi \partial_j \phi .
\ee
The scalar field can be expanded in terms of modes with Bunch-Davies initial condition as
\be
\phi = \int d^3 k \,\,\, e^{i\vec{k} \cdot \vec{x}} \left( \chi_k (\tau) a_{\vec{k}} + \chi^*_k (\tau) a^\dagger_{-\vec{k}} \right)
\ee
with $[a_{\vec{k}}, a^\dagger_{\vec{k}'} ] = \delta^3 (\vec{k} - \vec{k}' )$ and the modes $\chi_k$ satisfying\footnote{As is well-known, the equation of motion for the massless scalar is identical in form to that of the graviton presented earlier.} 
\be
(a\chi_k )'' + \left(k^2 - \frac{a''}{a} \right) (a\chi_k) = 0,
\ee
where the scale factor solves the classical slow-roll inflationary equations of motion and we have used primes to
denote derivatives with respect to conformal time. 
We compute the 4-point function of the scalar fields to be
\be
\label{scalar4pt}
\langle \partial_k \phi \partial_l \phi_1 \partial_i \phi_2 \partial_j \phi_2 \rangle = 
\int d^3p_1 d^3 p_2 e^{i(\vec{p}_1 + \vec{p}_2) (\vec{x}_1 - \vec{x}_2)}
\left[ p^k_1 p^i_1 p^l_2 p^j_2 + p^k_1 p^j_1 p^l_2 p^i_2 \right] 
 \chi_{p_1} (\tau_1 )\chi_{p_2} (\tau_1 ) \overline{\chi}_{p_1} (\tau_2 )
\overline{\chi}_{p_2} (\tau_2 )
\ee
where we have excluded tadpole diagrams. 
Contracting it with the graviton's polarization tensors yields
\be
\mathcal{G} (p_1, p_2, q) =  \frac{\beta}{p^3_1 p^3_2 }
 \left[ \left(p^2_1 - \frac{( \vec{p_1}\cdot \vec{q} )^2}{q^2} \right) \left(p^2_2 - \left( \frac{q^2+ \vec{p}_1 \cdot \vec{q} }{q} \right)^2 \right) \right],
\ee
where $\beta = \frac{2}{(2\pi)^3} \left(\frac{H}{M_p } \right)^4 \frac{1}{q^2}$ and $ \vec{p}_1 \cdot \vec{q} = \frac{1}{2} 
\left( p^2_2 - p^2_1 - q^2  \right)$. 
All time-dependent functions are captured in $F_1(p_1, p_2, q), F_2(p_1, p_2, q)$ which represent the two time integration contours. Unlike the case of the fermions, the scale factors do not explicitly cancel out as the scalar field is not conformally coupled. We proceed following \cite{Weinberg} where it was first explained how one could
nonetheless compute the integral with a good approximation even without a full solution to the slow-roll inflationary problem. The crucial point is that the integral has exponential terms which oscillate rapidly at early times ($p \gg aH)$ while at late times ( $p \ll a H$ ), the integral will converge exponentially fast and be dominated by the 
period surrounding horizon exit ($p \sim q \sim a H$). The assumption of slow-roll ($\epsilon \ll 1$) further implies that in computing the dominant contribution, we can take the mode functions to be  
\be
\chi_p (\tau) = \frac{H}{\sqrt{2} (2\pi )^{3/2} p^{3/2}} e^{-ip \tau} (1 + ip \tau ) + \mathcal{O} (\epsilon).
\ee
We can then proceed to perform the time integrals, with $a (\tau )\approx q/H (\tau_q)$ at each vertex. In \cite{Weinberg}, exactly the same procedure is used to obtain the one-loop correction term to the scalar curvature two-point function. For late-time correction terms, we refer the reader to an asymptotic expansion (in inverse powers of the scale factor) for the graviton and minimally coupled scalar mode wavefunctions in \cite{Weinberg} which is relevant in our context of slow-roll inflation where the scale factor expands much faster than the Hubble parameter $H$. In \cite{Senatore}, this identical approximation procedure was implemented for the loop correction due to the inflaton self-interactions in an effective theory framework.

In this case, we find
\bea
F_1 (p_1, p_2, q) &=& \int^{0}_{-\left(  \frac{\Lambda}{Hp_1} \right)_+} d\tau_2 \int^{\tau_2 (\frac{H}{\Lambda} + 1)}_{-\left(  \frac{\Lambda}{Hp_1} \right)_+} d\tau_1
\Bigg[ (1+iq \tau_1 )(1+ip_1 \tau_1 )(1+ip_2 \tau_1 )(1+iq\tau_2)(1-ip_1 \tau_2) (1-ip_2 \tau_2 ) \cr
&&\qquad \qquad \qquad \qquad  \times e^{-iq (\tau_1 + \tau_2) + i(p_1 + p_2)(\tau_2 - \tau_1)} \Bigg], \cr
F_2 (p_1, p_2, q) &=&     
 \int^{0}_{-\left(  \frac{\Lambda}{Hp_1} \right)_-} d\tau_2 \int^{0}_{-\left(  \frac{\Lambda}{Hp_1} \right)_+} d\tau_1
\,\,
 \Bigg[  (1 + i q\tau_1) (1-iq \tau_2) 
(1+ip_1 \tau_1)(1+ip_2 \tau_1) (1-ip_1 \tau_2) (1-ip_2 \tau_2) 
 \cr
\label{scalarTime}
&&\qquad \qquad \qquad \qquad \qquad \times e^{i(q + p_1 + p_2)(\tau_2 - \tau_1)} \Bigg]. 
\eea
In the similar vein as we have discussed the fermion case, in the early-time regime, the convergence of the time integrals is parametrized by $\Lambda$ as follows. In the large $\Lambda$ expansion, 
the leading-order term goes as 
$$
F_1 \sim \left(   
\frac{q^2 p^2_2}{(q+p_1+p_2)(p_1+p_2-q)p^4_1}
\right)
e^{-2\epsilon \Lambda \frac{q}{Hp_1}} \frac{\Lambda^6}{H^6} + \ldots +
\left(   
\frac{qp^2_2}{p^4_1  (q+p_1+p_2)}
\right)
e^{-2\epsilon \Lambda \frac{q}{Hp_1}} \frac{\Lambda^6}{H^6} + \ldots,
$$
where we have displayed the $\Lambda$-dependent leading-order terms corresponding to the lower and
upper limits of the $\tau_1$-integral in $F_1$ respectively. Similarly for $F_2$, the $\Lambda$-dependent leading-order term goes as
$$
F_2 \sim \left(
\frac{qp_2
(q^3+p^3_1+4qp^2_2 + p^3_2 + 4p^2_1 (q+p_2) + 4p_2 (q^2+3qp_2 + p^2_2)
}{p^2_1(q+p_1+p_2)^5}
\right)
e^{-\epsilon \Lambda  \frac{(q+p_1 + p_2)}{Hp_1}} \frac{\Lambda^3}{H^3} + \ldots
$$
These expressions show that the $\Lambda$-dependent integral limits of \eqref{scalarTime} do not enter
into the final expression of the logarithmic term we are seeking.

After completing the time integrals, we performed the momenta integrals in \eqref{iloop} to obtain in cutoff regularization
\be
I_1 + I_2 = \frac{78}{5} \text{Log} \tilde{L} + 4\left(
\frac{39}{10} + \frac{8}{\tilde{L}} - 10 \tilde{L} - 12 \tilde{L}^3 + \frac{34 \tilde{L}^5}{5}
\right) \text{Log} (1 + \tilde{L}^{-1} ) + \ldots,
\ee
where the ellipses are polynomials in $\tilde{L}$. And thus, after covariantizing the cutoff, we obtain the one-loop correction to be
\be
\label{Fscalar}
I_t = -\frac{1}{(2\pi )^2 q^3} \frac{H^4}{M^4_p}
 \times 
\left(
\frac{78}{5}
\right)
 \text{Log} \left( \frac{H}{\mu} \right). 
\ee
Contrary to the Dirac fermion case, this time we find that the second time integral contour does not contribute. 
(i.e. only $I_1$ in \eqref{iloop} contributes to \eqref{Fscalar}. 
From dimensional regularization, we also obtain an identical finite one-loop logarithmic correction term. 
Following \eqref{DimRegular}, each time integration contour gives 
\be
\frac{1}{\delta}  \int^\infty_1 dP \, J_1 (P, \delta ) 
 = \frac{2}{\delta}  \int^\infty_1 dP \, 
\frac{54}{5 P^{-\frac{4}{\delta}} } - 
\frac{1093}{105 P^{-\frac{2}{\delta}}   } 
+\frac{39}{5} -\frac{5P^{-\frac{1}{\delta}}}{3} + \ldots
\ee
and
\be
\frac{1}{\delta}  \int^\infty_1 dP \, J_2(P, \delta )  
=\frac{2}{\delta}  \int^\infty_1 dP \, 
=  \frac{5 P^{-\frac{1}{\delta}}  }{3} - \frac{81 P^{-\frac{3}{\delta}} }{28} +\ldots
\ee
Combining both $I_1$ and $I_2$, we obtain the finite term to be \eqref{Fscalar}. 
The overall sign of the logarithmic correction
is the same as that of fermions yet it arises from the different part of the in-in contour.

\subsection{The abelian gauge field}
\label{sec:Photon}

We consider an abelian gauge field residing on the FRW background with action
\be
S_{maxwell} = - \int d^4x \sqrt{-g}\frac{1}{4}g^{\mu \alpha} g^{\nu \beta} F_{\alpha \beta}
F_{\mu \nu}.
\ee
Keeping to first-order in metric perturbation, after performing a Legendre transform, we find the interaction Hamiltonian
\bea
H_{int} &=&\frac{1}{2} \int d^3x \,\,\,
 h_{\tau \nu} F^{\nu \mu} {F_{\mu}}^\tau \cr
&=& - \frac{1}{2} \int d^3x \,\,\,   h_{ik} \left(  \partial_0 A^k \partial_0 A^i - \partial^{[k} A^{j] }\partial^{ [ j }  A^{i ] } \right) 
\equiv \frac{1}{2} \int d^3x\,\,\, h_{ik} \mathcal{G}^{ki}.
\eea
The gauge field can be expanded in terms of modes as follows
\be
A_i (\vec{x}, \tau ) = \int d^3q\,\, \sum_\lambda e^{i\vec{q} \cdot \vec{x}} \left(  
e_i (\hat{q}, \lambda ) \,a_{\vec{q}, \lambda} \mathcal{A}_q (\tau) + e^*_i (-\hat{q}, \lambda ) \,a^\dagger_{-\vec{q}, \lambda} \mathcal{A}^*_q (\tau)
\right), \qquad A_0 = 0,
\ee
where $\lambda$ labels the polarization mode with the polarization vectors satisfying
$$
\sum_{\lambda, \lambda '} e^*_i (\hat{q}, \lambda ) e_j (\hat{q}, \lambda ' ) = \delta_{ij} - \hat{q}_i \hat{q}_j \equiv
P^{ij} (\hat{q} ),
$$ 
and the oscillators obeying the canonical relation $[a_{\vec{p}, \lambda } , a^\dagger_{\vec{p}', \lambda '}] = \delta^3 (\vec{p} - \vec{p}' ) \delta_{\lambda \lambda'}$.  Also, the mode wavefunctions are 
\be
\mathcal{A}_q (\tau ) = \frac{1}{(2\pi )^{3/2} \sqrt{2q}} e^{-iq \tau},
\ee
which solves Maxwell's equations on a conformally flat spacetime. We have gauge-fixed the system which
contains two ($\lambda = \pm 1$) physical propagating degrees of freedom. Substituting the mode expansion into the four-point function and omitting tadpole diagrams, we obtain after some algebra,
\bea
\langle 0 | \mathcal{G}^{lk} (x_1) \mathcal{G}^{ji} (x_2) | 0 \rangle &=&
\frac{1}{4} \int d^3p_1 d^3p_2 d^3p'_1 d^3p'_2
 \,\,\,  e^{i (\vec{p}_1 + \vec{p}_2 ) \cdot (\vec{x}_1 - \vec{x}_2 ) } \cr
&&
 \left[ \delta^3 ( \vec{p}_2 + \vec{p}'_1 ) \delta^3 (\vec{p}_1 + \vec{p}_2' ) + 
\delta^3 (\vec{p}_1 + \vec{p}'_1 ) \delta^3 (\vec{p}_2 + \vec{p}'_2 ) \right] 
\cr
&& \left( 
e^l_{p_1} e^k_{p_2} \dot{\mathcal{A}}_{p_1} (\tau_1 )  \dot{\mathcal{A}}_{p_2} (\tau_1 ) 
+ p^{[l}_1 e^{s]}_{p_1} p^{[s}_2 e^{k]}_{p_2} \mathcal{A}_{p_1} (\tau_1) \mathcal{A}_{p_2} (\tau_2 )
\right) \cr
&&\times 
\left( 
e^{*j}_{-p'_1} e^{*i}_{-p'_2} \dot{\mathcal{A}}^*_{p'_1} (\tau_2 )  \dot{\mathcal{A}}^*_{p'_2} (\tau_2 ) 
+ p'^{[j}_1 e^{*s]}_{-p'_1} p'^{[s}_2 e^{*i]}_{-p'_2} \mathcal{A}^*_{p'_1} (\tau_2) \mathcal{A}^*_{p'_2} (\tau_2 )
\right) 
\eea
where the delta functions arise from the VEV $\langle  0 | a_{\vec{p}_1} a_{\vec{p}_2} a^\dagger_{-\vec{p}'_1} a^\dagger_{-\vec{p}'_2} |0 \rangle$. It is convenient to express the virtual momenta in terms of unit vectors after which we are left
with a remnant dimension-2 factor of $p_1 p_2$. The mode wavefunctions all yield a common phase factor $e^{-i (p_1 + p_2) (\tau_1 - \tau_2)}$ and a constant
factor of $\frac{1}{4(2\pi )^6}$. Integrating over the primed momenta, we obtain 
\be 
\langle 0 | \mathcal{G}^{lk} (x_1) \mathcal{G}^{ji} (x_2) | 0 \rangle = \mathcal{N} \int d^3p_1 d^3p_2 \,\,\,  e^{i (\vec{p}_1 + \vec{p}_2 ) \cdot (\vec{x}_1 - \vec{x}_2 ) } 
e^{-i(p_1 + p_2) (\tau_1 - \tau_2 )} p_1 p_2 \mathcal{C}^{lkji} (p_1, p_2 )
\ee 
where 
$
\mathcal{N} = \frac{1}{2 (2\pi)^6} 
$
and $\mathcal{C}^{lkji} (p_1, p_2 )$ is a function of $p_1, p_2$ that reads
\bea
\mathcal{C}^{lkji} (p_1, p_2 ) &=& 
P^{li} (\hat{p}_1) {P}^{kj} (\hat{p}_2)
 - P^{l[i} (\hat{p}_1) \hat{p}^{s]}_1  P^{k[j} (\hat{p}_2) \hat{p}^{s]}_2 
- P^{i[l} (\hat{p}_1) \hat{p}^{s]}_1  P^{j[k} (\hat{p}_2) \hat{p}^{s]}_2 \cr
&+& \hat{p}^{[l}_1 \hat{p}^{s][i} (\hat{p}_1) \hat{p}^{r]}_1 
 \hat{p}^{[s}_2 \hat{p}^{k][r} (\hat{p}_2) \hat{p}^{j]}_2.
\eea
In obtaining $\mathcal{G}(p_1, p_2, q)$ in \eqref{iloop} , we find
the following identities useful:
\bea
P^{li} (\hat{p}_1) P^{li} (\hat{p}_2 ) &=& 1+(\hat{p}_1 \cdot \hat{q})^2, \cr
\hat{p}^l_1 \hat{p}^i_1  P^{li} (\hat{q}) &=& 1 - ( \hat{p}_1 \cdot \hat{q} )^2, \cr
P^{js} (\hat{p}_2) P^{si} (\hat{p}_1) &=& P^{ji} (\hat{p}_2) + \hat{p}^i_1 
\left( 
\hat{p}^j_2 (\hat{p}_2 \cdot \hat{p}_1) - \hat{p}^j_1
\right).
\eea
We find that contracting the spacetime indices with those of the graviton polarization tensors
yields (below, we introduce the symbols $S= \hat{p}_1 \cdot \hat{p}_2, 
R_1 = \hat{p}_1 \cdot \hat{q}, R_2 = \hat{p}_2 \cdot \hat{q}$ to simplify our notation)
\bea
\mathcal{G}(p_1, p_2, q)
&=& N \Bigg[ (1+S^2)(1+R^2_1)(1+R^2_2) 
+(1-R^2_1) (1-R^2_2)(1+S^2 ) \cr
&&+\, 2S\Bigg[
R^2_1 -1 + (1+S)(1-R^2_1 )^2 + (1-R^2_2 )[ R_1 R_2 - S  ] \cr
&&- (1-R^2_1)[
1+R^2_1 - (1+R_2)(1-S^2)
\Bigg]+2(R_2 -SR_1)(SR_2-R_1 ) \cr
&&-2S (1-S^2)+2(1-S^2)(1-R^2_2)(1+R^2_1) 
-4(1+S)(R^2_1 -1)S(1+R^2_2) \cr
&&- 4 (1-R^2_2) (R_2 - SR_1)SR_1 
-2S (1+R^2_1)(1+R^2_2) 
+4(R^2_1-1)(1+S)(1+R^2_2) \cr
&&-4R^2_1(R^2_1 -1)-2(1-R^4_1) -2(1-R^2_2)(1+S)(1-R^2_1) \Bigg]
\eea
where $N = \frac{2p_1 p_2}{(2\pi)^3 q^4} \frac{H^4}{M^4_p}$. 
We assemble all time-dependent parts of various modes wavefunctions into the
two time integrals which read
\bea
F_1 (p_1, p_2, q) &=&  \int^{0}_{-\left(  \frac{\Lambda}{Hp_1} \right)_+} d\tau_2 \int^{\tau_2 (\frac{H}{\Lambda} + 1)}_{-\left(  \frac{\Lambda}{Hp_1} \right)_+} d\tau_1\,\,
e^{-i q (\tau_1 + \tau_2 )}e^{-i (p_1 + p_2)(\tau_1 - \tau_2) }
(1+i q \tau_1 )(1+i q \tau_2 ) \cr
F_2 (p_1, p_2, q) &=&
\label{gaugeTime}
 \int^{0}_{-\left(  \frac{\Lambda}{Hp_1} \right)_-} d\tau_2 \int^{0}_{-\left(  \frac{\Lambda}{Hp_1} \right)_+} d\tau_1
\,\,e^{-i q (\tau_1 - \tau_2 )}e^{-i (p_1 + p_2)(\tau_1 - \tau_2) }
(1+i q  \tau_1 )(1- i q \tau_2 ) \nonumber \\
\eea
The time integrals are identical to those in the Dirac fermion case and we refer the reader to the discussion surrounding \eqref{fermionTime} for remarks on the early-time convergence parametrized by $\Lambda$. 
Upon performing the momenta integrals in \eqref{iloop}, we find the logarithmic divergence via cutoff regularization to be
\be
\label{Fgauge}
I_L =-\frac{1}{(2\pi)^2} \frac{H^4}{M^4_p} \frac{1}{q^3} 
\frac{2657}{315}
\text{Log} \left( \frac{H}{\mu} \right).
\ee
We find identical results from dimensional regularization. Following \eqref{DimRegular},
for $I_1$ in \eqref{iloop}, we have 
\be
\frac{1}{\delta}  \int^\infty_1 dP \, J_1 (P, \delta ) 
=2 \frac{1}{\delta} \int^\infty_1 dP \,\,
\left(
\frac{70}{3P^{-\frac{4}{\delta}}
} - \frac{14}{3  P^{-\frac{3}{\delta}}
} - \frac{233}{21  P^{-\frac{2}{\delta}}
} + \frac{493}{70P^{-\frac{1}{\delta}}
} + \frac{199}{126} - \frac{779 P^{-\frac{1}{\delta}}
}{315}
+ \ldots
\right),
\ee
whereas for $I_2$, we have
\be
\frac{1}{\delta}  \int^\infty_1 dP \, J_2(P, \delta )  
= 2\frac{1}{\delta} \int^\infty_1 dP \,\,
\left(
\frac{14}{3P^{-\frac{3}{\delta}}}- \frac{373}{70P^{-\frac{1}{\delta}}} + \frac{277}{105} + \frac{383P^{-\frac{1}{\delta}}}{315}
+ \ldots
\right).
\ee
Combining both $I_1$ and $I_2$, we obtain the finite term to be 
 -$\frac{2657}{315 \delta}$ after integrating and taking into account $q^{\delta}$ term, which leads to 
\eqref{Fgauge}

\section{On Seagull diagrams}
\label{sec:Seagull}

Seagull vertices (see Figure \ref{fig:f2})
can potentially contribute to the one-loop logarithmic correction term
since from \eqref{SK}, the first-order term is at one-loop order if $H_{int}$ is 
quadratic in $h_{ij}$,
\be
\label{sg}
I_L = - 2 \text{Im} \left(  \int^0_{-\infty_+} d\tau \langle H_{int} (\tau) h_{mn} (\vec{x},0) h_{mn} 
(\vec{x}', 0) \rangle \right).
\ee
On a closer inspection, we find that this cannot contribute to the 
one-loop logarithmic correction. In the following, we elaborate on this point. 

In \eqref{sg}, the correlation function factorizes into a graviton 4-point function
and a 2-point function of the matter fields. The two-point function can
be expressed as a momenta-integral, the integrand having two spacetime indices
(to be contracted with two other in the graviton's 4-point function) and of which 
scaling dimension reveals if it potentially harbors the logarithmic running. By 
assembling all terms quadratic in the graviton fields, one obtains the interaction Hamiltonian. 
After integrating over the dummy spatial coordinate and the virtual momenta of the
gravitons, one finds that for all cases, one ends up with the following 
schematic form for $I_L$:
\be
\label{seagull}
I_L =\int^0_{-\infty_+} d\tau\,\,
 \int d^3p\,\,\, \Phi^{I} (\vec{p}, \tau) \mathcal{H}^{I} (\vec{q}, \tau ),
\ee 
where $\Phi^{I} (\vec{p}, \tau),  \mathcal{H}^{I} (\vec{q}, \tau )$ are functions
arising from the matter and graviton correlation functions respectively, and 
we have used $I$ to denote a collective multi-index of which form depends on the 
interaction Hamiltonian. For all cases, the 2-point function of matter fields does not carry dependence of coordinates
and the 4-point function of graviton thus reduces to 
\bea
\label{seagullgraviton}
&&\int d^3x_1  \langle h_{kl} (x_1) h_{ij} (x_1) h_{mn} (\vec{x}, 0) h_{mn} ( \vec{x}', 0 ) \rangle =
8 \left( \frac{H}{M_p} \right)^4 e^{-2iq \tau} \frac{(1+iq \tau )^2}{q^6} \mathcal{P}^{klij} (\hat{q} ), \cr
&&\mathcal{P}^{klij} (\hat{q} ) \equiv  2 \left( 
P^{ki} (\hat{q}) P^{lj} (\hat{q}) + P^{kj} (\hat{q}) P^{li} (\hat{q}) - P^{kl} (\hat{q}) P^{ij} (\hat{q})
\right),
\eea
which does not enter into the one-loop correction. (For the fermions, we also have terms
involving derivatives of the graviton field which induce factors of the external momenta $q^m$
in \eqref{seagullgraviton}. )

In the following, we elaborate on the form of $\Phi^I (\vec{p}, \tau )$ for each type of matter field, explaining how
we deduce the absence of the one-loop correction term for each. The interaction Hamiltonian in \eqref{sg} can be obtained by expanding
\be
\label{int1}
\sqrt{-g} = \sqrt{-b} \left( 1 - \frac{1}{4} \gamma^\mu_\nu \gamma^\nu_\mu + \ldots \right),\,\,\,
\,\,g^{ij} = b^{ij} - \gamma^{ij} + \gamma^i_k \gamma^{kj} + \ldots, 
\ee
and for the fermions, we also have relevant terms second-order in $h_{ij}$ arising from the spin connection.
The scaling behavior of the $p-$integrand in \eqref{seagull} will reveal if there is any logarithmic correction immediately. Compared to the one-loop diagrams we evaluated earlier, we end up simply with a 3D momentum integral and look for terms of the form $1/p^3$ which if present will lead to the logarithmic running. 
The function $\mathcal{H}^{I} (\vec{q}, \tau)$ is independent of the internal momentum $p$ and hence its exact form is not required when we diagnosed the absence of the one-loop correction. 

We find that all seagull diagrams do not lead to finite one-loop correction, in agreement with a point first raised in \cite{Eugene}. Below, we present some details on the form of $\Phi^{I} (\vec{p}, \tau )$ for each matter field.
For the matter fields of integer spin, the only terms that could lead to \eqref{seagull} arise from
\eqref{int1}.

For the scalar field, the time integral induces some momentum-dependence since $\Phi^{ij}$ is time-dependent, 
with the relevant seagull vertices arising from 
\be
\label{scalarSG}
H_{int} \sim \int d^3y \, a^2  (\tau ) \left( 
 h_{ik} h_{kj} \partial_i \phi \partial_j \phi -  \frac{1}{4} h_{lk} h_{kl} \delta^{ij} \partial_i \phi \partial_j \phi \right),
\ee
with the scalar 2-point function being 
\be
\langle \partial_i \phi  \partial_j \phi \rangle = \int d^3p \,  \frac{(1+p^2 \tau^2 ) p^i p^j}{2(2\pi)^3 p}.
\ee
After performing the time-integral, we obtain 
\be
I =\left[  \int d^3p \frac{p^i p^j}{p} \frac{5iq^2 - 7ip^2}{4q^3} \right]\frac{4H^4}{M^4_p (2\pi)^3 q^6}
\left(
\frac{1}{4} \mathcal{P}^{bcbc} (\hat{q} ) \delta^{ij} - \mathcal{P}^{irjr} (\hat{q} )
\right),
\ee
from which we deduce that there is no one-loop correction term.

For the photon field, the seagull interaction Hamiltonian reads
\be
H_{int} = \int d^4x \,\, 
\left[ \frac{1}{4} h_{ij} h_{mn} F_{jn} F_{im} + \frac{1}{2} h_{ik} h_{kj} 
F_{\beta j} F_{\beta i} - 
\frac{1}{4} h_{ik} h_{ki} \left( -F_{\beta m} F_{\beta m} + F^{0m} F_{0m} \right)
-\frac{1}{2} h_{im} h_{in} F^{0m} F^{0n}
 \right].
\ee
Consider the VEVs $\langle F_{0m} F_{0j} \rangle, \langle F_{jn} F_{im} \rangle$. Using
the mode expansion of the gauge field, we find
\bea
\langle F_{jn} F_{im} \rangle &=& \frac{1}{2(2\pi)^3} \int d^3p \, p\, \hat{p}^{[j}   \delta^{n][m}  \hat{p}^{i]}, \cr
\langle F_{0m} F_{0j} \rangle &=& \frac{1}{2(2\pi)^3} \int d^3p \left(  \delta^{mn} - \hat{p}^n \hat{p}^m \right) p ,
\eea
which thus clearly indicates that the $p$-integral in \eqref{seagull} does not yield 
the logarithmic term.

For the fermions, apart from \eqref{int1} we need to take into account the spin connection. 
We find the second-order interaction Hamiltonian to read 
\bea
H_{int} &=& \int d^3x\, a^3 (\tau )\,
 \left(  \frac{1}{4} h_{ij} h_{ij} \delta_{m n}  - \frac{3i}{16} h_{m k} h_{k n} \right)
 \frac{i}{2} \left( \overline{\varphi} \gamma^m \partial^n \varphi
- \partial^n \overline{\varphi} \gamma^m \varphi \right) \cr
&-& \frac{i}{16} h_{n a} \partial^a h^{mn} \left( \overline{\varphi} \gamma_m \varphi \right)
- \frac{i}{32} \left( h^{s a} \partial^m h^n_a - h^{n a} \partial_m h^s_a 
\right) \left( 
\overline{\varphi} \gamma_m \gamma_n \gamma_s \varphi
\right).
 \eea
We consider the VEVs of the spinor field components which read
\bea
\langle \overline{\varphi}_a \varphi_b \rangle &=& \int d^3p\, \frac{\gamma^\mu_{ba} \hat{p}_\mu}{2(2\pi)^3 a^3 (\tau)},\cr
\langle \overline{\varphi}_a i\partial_m \varphi_b \rangle &=& \int d^3p\, \frac{ip_m \gamma^\nu_{ba} \hat{p}_\nu}{2(2\pi)^3 a^3 (\tau)}, 
\eea
from which we observe that the $p$-integral in \eqref{seagull} does not contain any one-loop logarithmic term.


\section{Discussion}
\label{sec:Discussion}

We have implemented dimensional regularization and cutoff regularization 
in the Schwinger-Keldysh formalism
following the broad framework first 
presented in \cite{Weinberg,Senatore} to study 
one-loop corrections to the two-point correlation function of tensor perturbations
in primordial cosmology induced by massless scalar, fermion and abelian gauge fields.
For all cases, we found a logarithmic running of the form
\be
\label{tensoroneloop}
\langle
h_{mn} h_{mn} \rangle_{1-loop} = \frac{C}{q^3} \frac{H^4}{M^4_p} \log \left( \frac{H}{\mu} \right), \,\,\, H = \frac{q}{a(\tau_q)},
\ee
and determined the constant $C$ for each isocurvature field which we found to be negative for all the massless fields considered.
Both regularization methods yield the same result in all cases which furnishes a consistency check of our computations. Our results should play an integral role in the
fully renormalized one-loop correction terms which we leave for future work. In this regard, it would be interesting to study how the
renormalization prescription in \cite{Mark2017,Mark2013} can be applied to our result 
\eqref{tensoroneloop1}.

Moving beyond the context of our work, we note that
recent explorations of the cosmological collider program 
(\cite{Maldacena}, \cite{Baumann}) have indicated that the mass and spin of particles present during inflation can leave their imprints on 
various cosmological correlation functions and loop effects play a certain role. Another natural future direction
would be to extend our calculation to gravitino fields \cite{Senatore2,Tan2021} and we have gathered some related preliminary calculations in the (unpublished) Appendix. More broadly speaking,
we hope that our discussion of the regularization procedures for the one-loop corrections of the tensor spectrum will be relevant and useful for
computing loop corrections in other applications of cosmological perturbation theory.

\section*{Acknowledgments}
I am very grateful to an anonymous referee for various useful advice, especially for his/her comments on the regularization procedures presented in \cite{Senatore} which led to an improvement over a previous manuscript draft.
I am indebted to Chu Chong-Sun, Neal Snyderman, Ori Ganor, Petr Horava and Daniel Robbins for their moral support during the course of completion of this work. This work is partially supported by a research fellowship at the School of Physical and Mathematical Sciences, Nanyang Technological University of Singapore. 

\begin{appendix}


\section{Some notes on the one-loop term for a massless spin-$\frac{3}{2}$ field}\footnote{This Appendix has
been excluded from the formal publication of this paper in JHEP. }
\label{sec:Gravitino}
In this (unpublished) Appendix, we furnish some aspects of a similar study of the one-loop term for 
a free, massless spin-$\frac{3}{2}$ field as described by the Rarita-Schwinger 
Lagrangian defined on the FRW background. Hopefully, this will be useful towards completing the computation of
the one-loop term for the gravitino on a slow-roll inflationary background, which requires a more careful consideration of other fields
as discussed in \cite{Senatore2}.

In SUGRA theories, the gravitino field
is typically described by a Majorana spinor field with a vector index $\psi^\mu$ that belongs to the 
$\left[ (\frac{1}{2}, 0) \oplus (0, \frac{1}{2} )\right] \otimes \left( \frac{1}{2}, \frac{1}{2} \right)$
representation of the Lorentz group.\footnote{See for example \cite{WeinbergQFT} for a nice review on this point.}
 One can extract the $\left( 1, \frac{1}{2} \right) \oplus \left( 
\frac{1}{2}, 1 \right)$ components by imposing a suitable gauge .

What is crucial
for our one-loop computation is an explicit expression for the spin-$\frac{3}{2}$ helicity sum which we develop in this section
and by which we demonstrate how each propagating degree of freedom can be neatly written as the product
of a spin-1 and spin-$\frac{1}{2}$ field components. After deriving the interaction Hamiltonian and the helicity sum
formula for the gravitino field, we then compute the one-loop correction just as in the preceding sections.

\subsection{Gravitino propagating on the FRW background and 
a helicity sum formula for the massless spin-$\frac{3}{2}$ field.}

We first recall that in Minkowski spacetime, the massive Rarita-Schwinger equation reads\footnote{See for example \cite{Moroi} and \cite{Giudice} for a nice exposition of the Rarita-Schwinger equation in relation to cosmology.}
\be
\label{RS}
\frac{i}{2} \left(  \gamma^\alpha \gamma^\mu \gamma^\beta - \gamma^\beta \gamma^\mu \gamma^\alpha \right) \partial_\beta \psi_\mu
+\frac{m}{2} [ \gamma^\alpha, \gamma^\beta ] \psi_\beta = 0,
\ee
where $\psi^\alpha$ is a set of four Majorana spinors. Consider first the massive gravitino setting where the gravitino field possesses both a spin-$\frac{3}{2}$ component and a spin-$\frac{1}{2}$ longitudinal mode. Contracting the LHS of  \eqref{RS} with $\gamma_\alpha$ and $\partial_\alpha$, we obtain
the equations
\be
\gamma^\alpha \psi_\alpha = \partial^\alpha \psi_\alpha = 0
\ee
which imply that we have two dynamical spinor fields and they correspond to spin-$\frac{3}{2}$ and spin-$\frac{1}{2}$
fields with each spinor satisfying the Dirac equation. 

In this paper, for simplicity, we focus on the massless case where we could find an explicit expression
for the propagator in Fourier space such that it can be employed for the one-loop computation. 
In the massless case, the equation of motion is invariant under
\be
 \psi^\alpha \rightarrow \psi^\alpha + \partial^\alpha \epsilon ,
\ee
 for some spacetime-dependent spinor $\epsilon$. This gauge symmetry can be partially fixed by choosing $\gamma^\alpha \psi_\alpha = 0$ (which
unlike the massive case does not necessarily follow from the equation of motion). The residual gauge symmetry is parametrized by some $\epsilon$
satisfying $\gamma^\alpha \partial_\alpha \epsilon =0$. We can thus impose one more constraint to fully fix the gauge ending up with 
only spin-$\frac{3}{2}$ fields capturing the physical degrees of freedom. In our paper, we adopt 
\be
\psi^0 = 0, \qquad \gamma^\alpha \psi_\alpha = 0,
\ee
to be our gauge conditions for the massless gravitino field (see also Section 3 of \cite{Kallosh:1999}). 

On the cosmological FRW background, much of the features discussed above remain the same. The partial derivative is replaced by a covariant one as follows
\be
D_\mu = \partial_\mu + \frac{1}{8} \omega_{\mu a b} \left[ \gamma^a, \gamma^b \right], \,\,\,
\omega_{\mu a b} = \frac{1}{2} \left( - C_{\mu a b} + C_{a b \mu } + C_{b \mu a} \right), \,\,\, {C^a}_{\mu \nu} = \partial_\mu e^a_\nu 
- \partial_\nu e^a_\mu - \frac{1}{2M^2_p} \overline{\psi}_\mu \gamma^a \psi_\nu,
\ee
where $\frac{1}{2M^2_p} \overline{\psi}_\mu \gamma^a \psi_\nu$ is the torsion term in the connection. 
The torsion could only modify the primordial tensor spectrum at three-loop level and 
for our purpose here we will not consider it further. 
Our starting point is the Rarita-Schwinger action 
\be
\label{RSaction}
\mathcal{S}_{rs} = i \int d^4 x \sqrt{-g} \frac{1}{3!} \overline{\psi}_\mu \Gamma^{\mu \nu \rho} \mathcal{D}_\nu \psi_\rho,
\ee
where $\Gamma^\mu = e^\mu_a \gamma^a$ and $\Gamma^{\mu \nu \rho} = \Gamma^{[\mu } \Gamma^\nu \Gamma^{\rho ] }$ 
is the completely anti-symmetrized product of the Dirac matrices. We note that in the de Sitter limit, 
\eqref{RSaction} is the gravitino's kinetic term in the action for pure de Sitter $\mathcal{N}=1$ SUGRA as formulated in \cite{Bergshoeff:2015} where it was shown that the de Sitter background breaks all supersymmetry.
\footnote{As presented in detail in \cite{Bergshoeff:2015}, the gauge-fixed $\mathcal{N}=1$ pure (no other matter fields apart from the gravitino multiplet)
de Sitter SUGRA has the Lagrangian of the form $$\mathcal{L} = \frac{1}{16\pi G} \left[ R - 
\overline{\psi}_\mu \Gamma^{\mu \nu \rho} \mathcal{D}_\nu \psi_\rho + \mathcal{L}_{\text{torsion}}
- \Lambda + \frac{m}{16\pi G} \overline{\psi}_\mu \Gamma^{\mu \nu} \psi_\nu \right],$$ where $R$ is the Ricci scalar. The de Sitter background
was shown in Section 4.2 of \cite{Bergshoeff:2015} to have no Killing spinors, and the SUSY-breaking scale
is set by $\Lambda + \frac{3m^2}{8\pi G}$. See also earlier works such as \cite{Giudice,Kallosh:1999,Waldron}
for various comments on the gravitino equations of motion and their massless $m=0$ limit, and Section 2.3 of \cite{Senatore2} for a broader EFT action. }

In the following, we will derive the gravitino's propagator in
Fourier space explicitly, consistent with a related computation in \cite{Bergshoeff:2015} where it was shown through the absence of zero modes in the Euclidean signature that
the gravitino wave operator (in the kinetic term) is invertible.

In the absence of torsion, we note that 
$
\mathcal{D}_0 = \partial_0, \,\,\, \mathcal{D}_k = \partial_k + \frac{H}{2} \gamma_k \gamma^0.
$
The equations of motion can be simply expressed as 
\be
\gamma^{\mu \nu \rho} \left( \partial_\nu + \Omega_\nu \right) \psi_\rho = 0.
\ee
We find that upon 
imposing the gauge conditions $\psi^0 = 0, \gamma^\alpha \psi_\alpha = 0$ and working in conformal time, 
the $\mu = 0$ equation is identically satisfied whereas taking $\mu $ to be a spatial index yields
\be
\label{RSfrw1}
\gamma^0 \partial_\tau \psi^{(T)}_k + \gamma^j \partial_j \psi^{(T)}_k + \frac{H}{2} \gamma^0 \psi^{(T)}_k = 0,
\ee
where we have used the superscript on $\psi^{(T)}$ to denote the fact that the spinor has been gauge-fixed. In particular
\eqref{RSfrw1} implies that if we define 
\be
\psi^{(T)}_k = \frac{1}{\sqrt{a}} \Psi_k,
\ee
then $\gamma^\mu \partial_\mu \Psi_k = 0$ so we can understand $\Psi_k$ to be satisfying the ordinary Dirac equation on flat 
spacetime.\footnote{Now the spacetime index on $\psi$ is raised/lowered using the metric tensor, so this implies that \eqref{RSfrw1}
can also be written as 
$
\gamma^0 \partial_\tau \psi^{(T)k} + \gamma^j \partial_j \psi^{(T) k} + \frac{5H}{2} \gamma^0 \psi^{(T) k} = 0,
$
which implies the identification
$
\psi^{(T) k} = \frac{1}{a^{5/2}} \Psi^k.
$
 }
We note that the gauge contraints can be expressed through a projection. In momentum space, we find that we 
can decompose the spinor into the gauge-fixed piece and remaining degrees of freedom as follows. 
\bea
\label{projection}
\psi_i &=& \psi^{(T)}_i + \left( \frac{1}{2} \gamma_i - \frac{1}{2} \hat{k}_i ( \hat{k} \cdot \vec{\gamma } ) \right) \gamma^j \psi_j - 
\left(
\frac{3}{2} \hat{k}_i + \frac{1}{2} \gamma_i ( \hat{k} \cdot \vec{\gamma } )
\right) \hat{k}^j \psi_j, \cr
\text{or}\,\,\,
\psi^{(T) i} &=& \left[ \delta^i_j - 
\left( \frac{1}{2} \gamma_i - \frac{1}{2} \hat{k}_i ( \hat{k} \cdot \vec{\gamma } ) \right) \gamma^j + 
\left(
\frac{3}{2} \hat{k}_i + \frac{1}{2} \gamma_i ( \hat{k} \cdot \vec{\gamma } )
\right) \hat{k}^j 
\right] \psi^j.
\eea
One can check that \eqref{projection} implies the gauge conditions $\gamma^\alpha \psi^{(T)}_\alpha = k^\alpha 
\psi^{(T)}_\alpha = 0$. Henceforth, we will be working with the gauge-fixed spinor and hence will drop the
superscript on the spinor to ease notations.

In momentum space, the fourier components of the projected spinors satisfy the following sum rule which is essential for our computation of the loop corrections later.
\be
\label{projector}
\mathcal{P}^{ij} (\vec{k} ) = \sum_{\lambda = \pm \frac{3}{2}} \psi^i(\vec{k}, \lambda ) \overline{\psi}^j (\vec{k}, \lambda ) = \frac{N_k}{2} \left(
\gamma^j - \hat{k}^j \left(\hat{k} \cdot \vec{\gamma} \right) 
\right) \gamma^\mu k_\mu \left( \gamma^i - \hat{k}^i \left(  \hat{k} \cdot \vec{\gamma} \right) \right),
\ee
where $\lambda$ labels the spin-$\frac{3}{2}$ helicity states, and $N_k = \frac{1}{2k (2\pi )^3 a(\tau) }$ is a normalization factor identical to the case of massless Dirac fermions on
a conformally flat spacetime. The projector satisfies the gauge contraints
$$
\gamma_i \mathcal{P}^{ij} = \hat{k}_i \mathcal{P}^{ij} = 0.
$$
In the ordinary Dirac fermion theory, the spinor field has no vectorial index and the analogue of \eqref{projector}
is simply the propagator. We will need \eqref{projector} in computing the correlation function of the spinor fields
in our calculation of the one-loop correction. 

In the following, we derive \eqref{projector} and in the process demonstrate how
each of the two spin-$\frac{3}{2}$ components of the projected gravitino spinor can be understood as arising from taking a suitable product of the massless spin-1 field polarization vector and a chiral spinor.\footnote{We note that our result is largely the same as an expression presented in \cite{Manoukian} which carried out a different derivation and of which final result differs from ours possibly due to spinor normalization among other reasons.} We will work in momentum
space and for the purpose of deriving \eqref{projector}, we need a specific representation of the Dirac matrices.
For convenience, we pick the Weyl representation in which
\be
\gamma^\mu = \left(    
\begin{array}{cc} 0 & \sigma^\mu \\
\overline{\sigma}^\mu & 0
\end{array}
\right), \qquad \sigma^\mu = (1, \vec{\sigma} ),\,\,\,\, \overline{\sigma}^\mu = (1, -\vec{\sigma} ).
\ee
A 4-dimensional Dirac spinor $U(\vec{k} )$ can be decomposed in terms of a pair of massless left- and right-handed spinors by writing
\be
U(\vec{k} ) = \left( \begin{array}{c} u_L (\vec{k} ) \\ u_R (\vec{k} ) \end{array} \right) \equiv U_L (\vec{k} ) \oplus
U_R (\vec{k} ),
\ee
where we normalize the two-component spinors as follows
\be
\label{chiralproj}
u_R (\vec{k} ) u^\dagger_R (\vec{k} ) = N_k k_\mu \overline{\sigma}^\mu, 
\qquad
u_L (\vec{k} ) u^\dagger_L (\vec{k} ) = N_k k_\mu \sigma^\mu.
\ee
On the other hand, it is known that the massless spin-1 polarization vectors can be expressed in terms of these
chiral spinors as follows (see for example \cite{Kosower} ). 
\be
\label{heliVec}
e^i_{-1} (\vec{k} ) = - \frac{1}{\sqrt{2}} \frac{
\overline{U}_R (- \vec{k} ) \gamma^i U_R (\vec{k} )
}{
\overline{U}_R (-\vec{k} ) U_L (\vec{k} )
}, \,\,\,\,\,
e^i_{1} (\vec{k} ) = - \frac{1}{\sqrt{2}} \frac{
\overline{U}_L (- \vec{k} ) \gamma^i U_L (\vec{k} )
}{
\overline{U}_L (-\vec{k} ) U_R (\vec{k} )
}.
\ee
Substituting \eqref{chiralproj} into
\eqref{heliVec} yields the following useful relations for the spin-1 polarization vectors.
\bea
e^i_{-1} (\vec{k} ) e^{*j}_{-1} (\vec{k} ) &=&     
\frac{U^\dagger_R (-\vec{k} ) \gamma^i U_R (\vec{k} ) U^\dagger_R (\vec{k} ) \gamma^j U_R (-\vec{k} )}
{2 U^\dagger_R (-\vec{k} ) U_R (\vec{k} ) U^\dagger_R (\vec{k} ) U_R (-\vec{k} )}
\cr
&=& \frac{1}{8} \text{Tr} \left(  
\sigma^i u_R (\vec{k} ) u^\dagger_R (\vec{k} ) \sigma^j u_R (-\vec{k} ) u^\dagger_R (-\vec{k} )
\right) \cr
\label{vectordown}
&=& \frac{1}{2} \left( 
\delta^{ij} - \hat{k}^i \hat{k}^j + i \epsilon^{ijm} \hat{k}_m
\right) 
\equiv \frac{1}{2}
P^{ij}_{\downarrow} (\hat{k} ),
\eea
and similarly, 
\be
\label{vectorup}
e^i_{1} (\vec{k} ) e^{*j}_{1} (\vec{k} ) =
\frac{1}{2} \left( 
\delta^{ij} - \hat{k}^i \hat{k}^j - i \epsilon^{ijm} \hat{k}_m
\right) \equiv \frac{1}{2}
P^{ij}_{\uparrow} (\hat{k} ).
\ee
We note that \eqref{vectordown} and \eqref{vectorup} imply that 
$
\sum_{\lambda = \{-1,1\} } e^i_\lambda e^{*j}_\lambda =   
\delta^{ij} - \hat{k}^i \hat{k}^j = P_{ij} =  \frac{1}{2} \left(  
P^{ij}_{\downarrow} (\hat{k} ) + P^{ij}_{\uparrow} (\hat{k} )
\right).
$
In the Weyl representation, we find that \eqref{projector} can be written as 
\be
\label{projector2}
\mathcal{P}^{ij} = 
\left( 
\begin{array}{cc} 0 & u_L u^\dagger_L e^i_{1}  e^{*j}_{1} \\
u_R u^\dagger_R e^i_{-1}  e^{*j}_{-1}  & 0
\end{array}
\right)
= 
e^i_{-1} (\vec{k} ) e^{*j}_{-1} (\vec{k} ) U_R \overline{U}_R + 
e^i_{1} (\vec{k} ) e^{*j}_{1} (\vec{k} ) U_L \overline{U}_L .
\ee
Comparing \eqref{projector2} with \eqref{projector}, we see that 
the massless spin-$\frac{3}{2}$ field has components that can be simply
decomposed in terms of massless spin-1 and spin-$\frac{1}{2}$ degrees of freedom, i.e. we can identify
\be
\psi^i \left(\vec{k}, \frac{3}{2} \right) = e^i_1  (\vec{k} ) U_L (\vec{k} ), \qquad \psi^i \left(\vec{k}, -\frac{3}{2} \right) = e^i_{-1} (\vec{k} ) U_R (\vec{k} ), 
\ee
where $U_{L,R} (\vec{k} )$ are the left- and right-handed massless Dirac spinors of spin $\pm \frac{1}{2}$ respectively. This completes
our derivation of the helicity sum formula \eqref{projector} which we need for calculating our one-loop correction term.

\subsection{The interaction Hamiltonian from linear fluctuations}
In this Section, we derive the interaction Hamiltonian for the massless gravitino. Our final result is the following expression:
\be
\label{intHamiltonianRS1}
H_{int}  = i \int d^3x  \, a(\tau )\,\,\,
h^{ik} \left[ 
\overline{\psi}_i \gamma^r \partial_r \psi_k + \frac{1}{2} \left(  
\overline{\psi}_j \eta^{jm} \gamma_k \partial_i \psi_m + \overline{\psi}_j \gamma_i \partial^j \psi_k 
\right)
\right],
\ee
where in our notational conventions, the indices (apart from those on the spinors) in \eqref{intHamiltonianRS1} are raised/lowered via Minkowski metric .  In the following, we present the outline for the derivation of 
\eqref{intHamiltonianRS1}. 

Up to first order, the spin connection remains unchanged and we only need to consider the linear fluctuations of the vielbeins. From
\be
\Gamma^{\mu \nu \rho} \mathcal{D}_\nu \psi_\rho = e^\mu_a e^\nu_b e^\rho_c \gamma^{abc} \left( \partial_\nu + \Omega_\nu \right) \varphi_\rho,
\qquad 
e^\mu_a = \frac{1}{2} \left[ \delta^\mu_a - \frac{1}{2}h^\mu_a + \ldots \right]
\ee
we obtain the interaction Lagrangian to be 
\be
\label{Lint}
 \mathcal{L}_{int} = i\int d^3x \,a(\tau )\, \overline{\psi}_\mu \mathcal{P}^{\mu \nu \rho}_{abc} \gamma^{abc} \mathcal{D}_\nu \psi_\rho,
\ee
where 
\be
 \mathcal{P}^{\mu \nu \rho}_{abc} = -\frac{1}{2} \left( 
\delta^\mu_a \delta^\rho_c h^\nu_b + \delta^\nu_b \delta^\rho_c h^\mu_a + \delta^\mu_a \delta^\nu_b h^\rho_c 
\right).
\ee
Imposing $\psi^0=0$, \eqref{Lint} then becomes
\be
\label{Lint2}
\mathcal{L}_{int} = i \int d^3x \, a (\tau )\,
 \overline{\psi}_i \left(
\mathcal{P}^{i0k}_{abc} \gamma^{abc} \partial_0 \psi_k + \mathcal{P}^{ijk}_{abc} \gamma^{abc} \left( \partial_j + \Omega_j \right) \psi_k
\right).
\ee
In the following, we furnish some details as to how we obtain \eqref{intHamiltonianRS1} from \eqref{Lint2}: 
\begin{enumerate}[(i)]
\item 
We find that the term involving the spin connection $\Omega$ in \eqref{Lint2} vanishes. This term is 
\be
\label{connectionTerm}
\overline{\psi}_i \mathcal{P}^{ijk}_{abc} \gamma^{abc} \Omega_j \psi_k = 
-\frac{1}{2} \overline{\psi}_i  \left(  
\gamma^{ibk} h^j_b + \gamma^{bjk} h^i_b + \gamma^{ijb} h^k_b 
\right) \frac{H}{2} \gamma_j \gamma^0 \psi_k
\ee
For each term in \eqref{connectionTerm}, we find
\bea
\overline{\psi}_i \gamma^{ibk} h^j_b \gamma_j \gamma^0 \psi_k&=& -12 h^k_b \eta^{bi} \overline{\psi}_i \gamma^0 \psi_k, \cr
\overline{\psi}_i \gamma^{bjk} h^i_b \gamma_j \gamma^0 \psi_k&=&  6 h^i_b \eta^{kb} \overline{\psi}_i \gamma^0 \psi_k, \cr
\overline{\psi}_i \gamma^{ijb} h^k_b \gamma_j \gamma^0 \psi_k&=& 6 h^k_b \eta^{bi} \overline{\psi}_i \gamma^0 \psi_k,
\eea
and thus $\overline{\psi}_i \mathcal{P}^{ijk}_{abc} \gamma^{abc} \Omega_j \psi_k = 0$.

\item For the remaining terms involving derivatives of $\psi$, we find 
\bea
\overline{\psi}_i \gamma^{a0k} h^i_a \partial_0 \psi_k &=& \overline{\psi}_i \gamma^{i0c} h^k_c \partial_\tau \psi_k
= 6 h^i_a \eta^{ak} \overline{\psi}_i \gamma^0 \partial_\tau \psi_k, \cr
\overline{\psi}_i \gamma^{ibk} h^j_b \partial_j \psi_k &=& 6 h^j_b \eta^{ik} \overline{\psi}_i \gamma^b \partial_j \psi_k, \cr
\overline{\psi}_i \gamma^{ijb} h^k_b \partial_j \psi_k &=& 6 \eta^{bi} h^k_b \overline{\psi}_i \gamma^j \partial_j \psi_k - 6 \eta^{ij} h^k_b \overline{\psi}_i \gamma^b \partial_j \psi_k.
\eea
\end{enumerate}
Assembling all the terms together and after a Legendre transformation, we then obtain \eqref{intHamiltonianRS1}. 
\footnote{We note in passing that we can make the reality condition of the classical Hamiltonian real by writing 
$H_{int} (\vec{x}, \tau ) = \frac{1}{2} 
a(\tau) h^{ik} \Bigg[ 
\overline{\psi}_i \gamma^r  \partial_r \psi_k  - \partial_r \overline{\psi}_i \gamma^r  \partial_r \psi_k
+ \frac{1}{2} \Bigg(  
\overline{\psi}_j \eta^{jm} \gamma_k \partial_i \psi_m  - \partial_i \overline{\psi}_j \eta^{jm} \gamma_k  \psi_m 
+ \overline{\psi}_j \gamma_i \partial^j \psi_k  - \partial^j \overline{\psi}_k \gamma_i  \psi_j
\Bigg)
\Bigg].$
One can check that this yields the same results as adopting \eqref{intHamiltonianRS1}. }
We now proceed to compute the one-loop correction to the tensor spectrum.
We can write its mode expansion as
\be
\label{gravitino}
\psi_i (\vec{x}, \tau ) = \int d^3p \sum_{\lambda} e^{i \vec{p} \cdot \vec{x}} \psi_i (\vec{p}, \lambda ) e^{-ip \tau} a_{\vec{p}, \lambda} + 
e^{-i \vec{p} \cdot \vec{x}} \psi^c_i (\vec{p}, \lambda ) e^{ip \tau} a^\dagger_{\vec{p}, \lambda}
\ee
of which form ensures that $\psi^c_i (\vec{x}, \tau ) = \psi_i (\vec{x}, \tau )$, with the superscript `c' denoting charge conjugation
and the fermionic oscillators obey the anti-commutation relation 
$$
\{ a_{\vec{p}, \lambda}, 
a^\dagger_{\vec{k}, \lambda'} \} = \delta^3 (\vec{p} - \vec{k} ) \delta_{\lambda \lambda'}.
$$ 
From \eqref{intHamiltonianRS1}, 
the interaction Hamiltonian 
density is of the form 
$$
\overline{\psi}_m (x) {\mathcal{D}^{mnr}}_{kl} (x) \gamma^r  \psi_n (x), 
$$
where we define
$$
{\mathcal{D}^{mn}}_{klr} \gamma^r \equiv \delta^m_k \delta^n_l \gamma^r \partial_r - \frac{1}{2} \left( \delta^{mn} \gamma_l \partial_k + \delta^{mr} \delta^n_l \gamma_k \partial_r \right).
$$
In the absence of any differential/matrix-valued operator, the 4-point function involving $\psi, \overline{\psi}$ reads
\bea
\langle \overline{\psi}_l \psi_k \overline{\psi}_i \psi_j \rangle &=& 
\sum_{\lambda_1, \lambda_2 } \int d^3 p_1 d^3 p_2 
e^{-i(p_1 + p_2) (\tau_1 - \tau_2 )}
e^{i (\vec{p}_1 + \vec{p}_2 ) \cdot (\vec{x}_1 - \vec{x}_2 )} \cr
\label{corr4}
&&\times
 \overline{\psi^c}_l (\vec{p}_1 , \lambda_1) \psi_k (\vec{p}_2, \lambda_2 ) 
\Bigg[
\overline{\psi}_i (\vec{p}_2, \lambda_2 ) 
\psi^c_j (\vec{p}_1, \lambda_1 )
-
\overline{\psi}_i (\vec{p}_1, \lambda_1 ) 
\psi^c_j (\vec{p}_2, \lambda_2 )
\Bigg],
\eea
where we note that the first and second term in the bracket corresponds to the contraction between the second and third spinor field and that between the second and last spinor field respectively, the negative sign due to the anti-commuting nature of the oscillators. 

To proceed, we commit to various specific representations of the Dirac algebra at various points in the following computation. 
Earlier on, we had worked in the Weyl representation when demonstrating how the form of $\mathcal{P}^{ij}$
can be derived from decomposing a spin-$\frac{3}{2}$ field in terms of lower spin ones. For the purpose of computing the VEV, we now pick 
the Majorana representation\footnote{In this representation, we have $\gamma^0$ being Hermitian and $\gamma^k$ being anti-
Hermitian. For example, $\gamma^0 = \sigma^2 \otimes \sigma^1, \gamma^2 = \sigma^2 \otimes i\sigma^2, \gamma^1 = i\sigma^1 \otimes \mathds{1}, \gamma^3 = i \sigma^3 \times \mathds{1}$. But we won't actually need any particular choice
in the rest of our workings.  }
in which all the Dirac matrices are purely imaginary and 
\be
\varphi^c = \varphi^*.
\ee
Let us proceed to simplify the gravitino correlation function.The interaction Hamiltonian contains terms of the following
form
$$
\langle \overline{\psi}_l \gamma^a  \psi_k \overline{\psi}_i \gamma^b \psi_j \rangle.
$$
Using the anticommuting nature of the oscillators, the first term 
$
 \overline{\psi^c}_l (\vec{p}_1 , \lambda_1)\gamma^a  \psi_k (\vec{p}_2, \lambda_2 ) 
\overline{\psi}_i (\vec{p}_2, \lambda_2 ) \gamma^b
\psi^c_j (\vec{p}_1, \lambda_1 )           $
can be easily seen to be equivalent to 
$$
\text{Tr} \left[ \mathcal{P}^{jl} (\vec{p}_1 ) \gamma^a \mathcal{P}^{ki} (\vec{p}_2) \gamma^b \right].
$$
The second term $
 \overline{\psi^c}_l (\vec{p}_1 , \lambda_1) \gamma^a \psi_k (\vec{p}_2, \lambda_2 ) 
\overline{\psi}_i (\vec{p}_1, \lambda_1 ) \gamma^b
\psi^c_j (\vec{p}_2, \lambda_2 )$
is not so immediately obvious in terms of how it can be expressed in terms of the
projector $\mathcal{P}^{ij}$. 
Making visible the matrix component indices (with capitalized Roman indices), 
we find that we can massage it to be in the form 
\bea
&&\left( \gamma^0  \right)_{MN} \mathcal{P}^{li}_{MS} (\vec{p}_1) 
\left( \gamma^a \right)_{NJ} \mathcal{P}^{kj}_{JQ} (\vec{p}_2) \left( \gamma^b \right)_{SK} \left(  \gamma^0 \right)_{QK} \cr
&=& \text{Tr} \left( \gamma^a \mathcal{P}^{kj} (\vec{p}_2) \gamma^0 \left( \gamma^b \right)^T \mathcal{P}^{li (T)} (\vec{p}_1) \gamma^0 \right) \cr
&=& - \text{Tr} \left(  
\gamma^a \mathcal{P}^{kj} (\vec{p}_2) \gamma^b \mathcal{P}^{il} (\vec{p}_1)
\right),
\eea
where we have invoked the useful relation 
\be
\label{transposeProj}
\mathcal{P}^{ij (T) } (\vec{p} ) = - \mathcal{P}^{ji} (-\vec{p} ), \qquad 
\gamma^0 \mathcal{P}^{ij} (\vec{p} ) \gamma^0 = \mathcal{P}^{ij} (-\vec{p} ). 
\ee
So this implies that the second term is related to the first one by switching $i \leftrightarrow j$. 
But when differential operators are involved, we need to switch the momenta sign for the last two spinors. Taking into account
the full Hamiltonian, after some algebra,
we find the gravitino correlation function to be 
\bea
&&\int \int d^3p_1 d^3p_2 e^{-i (p_1 + p_2) (\tau_1 - \tau_2 ) + i (\vec{p}_1 + \vec{p}_2)\cdot (\vec{x}_1 - \vec{x}_2)} (p^s_1 + p^s_2 ) \cr
&& \times\text{Tr} \Bigg[
\left( \gamma^s 
 \mathcal{P}^{jl} (\vec{p}_1) \gamma^r \mathcal{P}^{ki} (\vec{p}_2) 
\right) p^r_2 (p^s_1 + p^s_2 ) + 
\left( \gamma^s  \mathcal{P}^{jb} (\vec{p}_1) \gamma^l \mathcal{P}^{ai} (\vec{p}_2) 
\right) p^k_2 (p^s_1 + p^s_2 ) \eta^{ab}
\cr
&+&  
\left( \gamma^s 
 \mathcal{P}^{jr} (\vec{p}_1) \gamma^k \mathcal{P}^{li} (\vec{p}_2) 
\right) p^r_2 (p^s_1 + p^s_2 ) + 
\frac{1}{2}
\left( \gamma^j  \mathcal{P}^{db} (\vec{p}_1) \gamma^l \mathcal{P}^{ac} (\vec{p}_2) 
\right) p^k_2 (p^i_1 + p^i_2 ) \eta^{ab} \eta^{cd} \cr
&+&  
\frac{1}{2} \left( \gamma^j 
 \mathcal{P}^{dr} (\vec{p}_1) \gamma^k \mathcal{P}^{lc} (\vec{p}_2) 
\right) p^r_2 (p^i_1 + p^i_2 ) \eta^{cd} 
+ \frac{1}{4}
\left( \gamma^i  \mathcal{P}^{jr} (\vec{p}_1) \gamma^k \mathcal{P}^{ls} (\vec{p}_2) 
\right) p^r_2 p^s_1  \cr
&+& 
\label{hamiltonianmain}
 \frac{1}{4}
\left( \gamma^i  \mathcal{P}^{sr} (\vec{p}_1) \gamma^k \mathcal{P}^{lj} (\vec{p}_2) 
\right) p^r_2 p^s_2  
\Bigg],
\eea
where we have in the process of simplification used \eqref{transposeProj} repetitively and the fact that we are integrating over
all momenta space of $p_1, p_2$ which can be interchanged as dummy variables. We also used the fact
that switching the sign of the momenta doesn't matter since it is multiplied to the graviton 4-point function which eventually depends on $\vec{q} = -\vec{p}_1 -\vec{p}_2$. At this point, we judiciously turn to the Weyl representation for an easier computation of the trace. \footnote{ Otherwise, without further 
simplification/insight, the algebra involves manipulating the spacetime indices contained in the trace of 8 gamma matrices which involve 105 terms involving products of kronecker delta tensor.}
In terms of Pauli matrices and the spin-1 projectors, each matrix trace reads 
\be
\label{matrixtrace}
\text{Tr} \left[  \gamma^a \mathcal{P}^{bs} (\vec{p_1} ) \gamma^r \mathcal{P}^{ij} (\vec{p_2} ) \right] = 
\text{Tr} \left( \sigma^a \overline{\slashed{p_1}} \sigma^r \overline{\slashed{p_2}}\right) P^{bs}_\downarrow (\hat{p_1} )P^{ij}_\downarrow 
(\hat{p_2} )
+ \text{Tr} \left( \sigma^a \slashed{p_1} \sigma^r  \slashed{p_2}\right) P^{bs}_\uparrow (\hat{p_1} )P^{ij}_\uparrow (\hat{p_2} )
\ee
with
\be
\text{Tr} \left[  \sigma^a \slashed{p_1} \sigma^r \slashed{p_2} \right] = 2 \delta^{ar} (p_1p_2 - \vec{p_1} \cdot \vec{p_2} ) + 
2p_1^a p_2^r + 2p_1^rp_2^a + 2i p_1p_2 \epsilon^{alr} (\hat{p_1}_l - \hat{p_2}_l ).
\ee
Defining 
$$
M^{ar} = \delta^{ar} (1 - \hat{p_1} \cdot \hat{p_2} ) + \hat{p_1}^r \hat{p_2}^a + \hat{p_2}^r \hat{p_1}^a + i \epsilon^{arl} ( \hat{p_1}^l - \hat{p_2}^l ),
$$
after some algebra, we find that the product of the gravitino and graviton correlation functions reads
\bea
&&\int \int d^3p_1 d^3p_2 e^{-i (p_1 + p_2) (\tau_1 - \tau_2 ) + i (\vec{p}_1 + \vec{p}_2)\cdot (\vec{x}_1 - \vec{x}_2)} (p^s_1 + p^s_2 ) \cr
&& \times \Bigg(  \Bigg[ 
2 M^{sr} P_{\uparrow}^{ki} (\vec{p}_2) P_{\uparrow}^{jl} (\vec{p}_1) q^r q^s 
+ 2 M^{sk} P_{\uparrow}^{jr} (\vec{p}_1) P_{\uparrow}^{li} (\vec{p}_2)  q^r q^s \cr
&&  +  M^{ik} P_{\uparrow}^{jr} (\vec{p}_1) P_{\uparrow}^{ls} (\vec{p}_2)  q^r q^s + 
+ M^{ik} P_{\uparrow}^{sr} (\vec{p}_1) P_{\uparrow}^{lj} (\vec{p}_2) q^r q^s \Bigg] \cr
&&
+ 2 M^{sl} P_{\uparrow}^{jb} (\vec{p}_1) P_{\uparrow}^{ai} (\vec{p}_2) \eta^{ab} q^k q^s + 
 M^{jl} P_{\uparrow}^{db} (\vec{p}_1) P_{\uparrow}^{ac} (\vec{p}_2) \eta^{ab} q^k q^i \cr
&&+  M^{jk} P_{\uparrow}^{dr} (\vec{p}_1) P_{\uparrow}^{lc} (\vec{p}_2) \eta^{cd} q^r q^i \Bigg) \times  \langle h^{kl} h^{ij} h_{mn} h_{mn} \rangle.
\eea
The integrand terms outside the square bracket can be shown to vanish since they involve momenta 4-vectors with the free indices
and they are multiplied to the graviton correlation function which contains polarization tensors satisfying
\be
\label{polF}
\sum_{\lambda, \lambda'} \epsilon_{kl} (\hat{q}, \lambda) \epsilon^*_{mn} (\hat{q}, \lambda ) \epsilon^*_{ij} 
(\hat{q}, \lambda' ) \epsilon_{mn} (\hat{q}, \lambda' ) 
= 2 \left( 
P^{ki} (\hat{q}) P^{lj} (\hat{q}) + P^{kj} (\hat{q}) P^{li} (\hat{q}) - P^{kl} (\hat{q}) P^{ij} (\hat{q}).
\right)
\ee
After using \eqref{polF} in the midst of some heavy algebra, we find that the contraction between
the gravitino and graviton correlation function yields the following function $\mathcal{G} (p_1, p_2, q)$ in 
the master formula \eqref{iloop}: 
\bea
\mathcal{G}(p_1, p_2, q) &=& N_G \Bigg( \left[  
1- \hat{p}_1 \cdot \hat{p}_2 + 2\hat{q} \cdot \hat{p}_1 \hat{q} \cdot \hat{p}_2 \right]
\left[ 
(1+  (\hat{q} \cdot \hat{p}_1 )^2)(1 + ( \hat{q} \cdot \hat{p}_2 )^2 )- 4 \hat{q} \cdot \hat{p}_2 \hat{q} \cdot \hat{p}_1
\right]
\cr
&&
+ 2\left(  (\hat{p}_1 \cdot \hat{p}_2)^2 -1 \right) - \hat{q} \cdot \hat{p}_1 \hat{q} \cdot \hat{p}_2 - 
(\hat{q} \cdot \hat{p}_1)^2 \left[  \hat{p}_1 \cdot \hat{p}_2 - 2 \hat{q} \cdot \hat{p}_1 \hat{q} \cdot \hat{p}_2 \right]
\cr
&&
-2 (1 - \hat{p}_1 \cdot \hat{p}_2 ) + 2 (\hat{q} \cdot \hat{p}_2 \hat{p}_1 \cdot \hat{p}_2 )(
1+3 \hat{q} \cdot \hat{p}_1 \hat{q} \cdot \hat{p}_2 ) -
6 (\hat{p}_1 \cdot \hat{p}_2 )^2 \hat{q} \cdot \hat{p}_1 - 4 \hat{q} \cdot \hat{p}_1 ( 
(\hat{q} \cdot \hat{p}_1 )^2 -1)
\cr
&&
+(1-\hat{q} \cdot \hat{p}_1 \hat{q} \cdot \hat{p}_2 )(1+  (\hat{q} \cdot \hat{p}_2)^2 )(1-(\hat{q} \cdot \hat{p}_1)^2)
-2 \hat{q} \cdot \hat{p}_2 ( \hat{q} \cdot \hat{p}_1 - \hat{q} \cdot \hat{p}_2 ) \cr
&&+2 \left(  (\hat{p}_1 \cdot \hat{q})^2 (\hat{q} \cdot \hat{p}_2 ) - \hat{q} \cdot \hat{p}_1 \hat{p}_1 \cdot \hat{p}_2               \right) (\hat{q} \cdot \hat{p}_1 - \hat{q} \cdot \hat{p}_2 )
\cr
&&
+(1 - \hat{q} \cdot \hat{p}_1 \hat{q} \cdot \hat{p}_2 ) 
\left[
(\hat{p}_1 \cdot \hat{p}_2 - \hat{q} \cdot \hat{p}_1 \hat{q} \cdot \hat{p}_2 )(  \hat{q} \cdot \hat{p}_1 \hat{q} \cdot \hat{p}_2  -1 ) + (1 - (\hat{q} \cdot \hat{p}_1 )^2 )(1 + (\hat{q} \cdot \hat{p}_2 )^2 )
\right]\Bigg), \nonumber \\
\eea
where
$
N_G = - \frac{1}{(2\pi)^3 q^4} \frac{H^4}{M^4_p} 
$
takes into account all the normalization constants. For the time integrals, from the gravitino correlation function, we have the factor
$
e^{-i (p_1 +p_2) (\tau_1 - \tau_2 )}
$
just like the case of the gauge field and fermions. Hence the time integrals $F_1(p_1, p_2, q)$ and $F_2(p_1, p_2, q)$
are the same as in those cases:
\bea
F_1(p_1, p_2, q) &=& \int^0_{-\infty_+} d\tau_2 \int^{\tau_2}_{-\infty_+} d\tau_1 e^{-i(p_1 + p_2)(\tau_1 - \tau_2)} 
(1+iq \tau_1 )(1+i q \tau_2 ) e^{-iq (\tau_1 + \tau_2)}, \cr
F_2(p_1, p_2, q) &=& \int^0_{-\infty_+} d\tau_1 \int^{0}_{-\infty_-} d\tau_2 e^{-i(p_1 +p_2)(\tau_1 - \tau_2)} 
(1+iq \tau_1 )(1-i q \tau_2 ) e^{-iq (\tau_1 - \tau_2)}.
\eea
Performing the momenta integral yields the logarithmic one-loop correction 
\be
\label{Fgravitino}
I_{gravitino} =- \frac{1}{(2\pi)^2} \frac{H^4}{M^4_p} \frac{1}{q^3} 
\frac{5107}{315}
\text{Log} \left( \frac{H}{\mu} \right),
\ee
a result which we also obtain from dimensional regularization. 
Following \eqref{DimRegular}, we find 
for $I_1$ in \eqref{iloop}
\be
\frac{1}{\delta}  \int^\infty_1 dP \, J_1 (P, \delta ) 
 = \frac{1}{\delta}  \int^\infty_1 dP \, 
\, \frac{218}{21P^{-\frac{2}{\delta}}
} + \frac{17}{70P^{-\frac{1}{\delta}}
} + \frac{5653}{630} + \ldots
\ee
whereas for $I_2$, we have 
\be
\frac{1}{\delta}  \int^\infty_1 dP \, J_2(P, \delta )  
 = -\frac{1}{\delta} \int^\infty_1 dP \,\,
\left( 
\frac{218}{105 P^{-\frac{1}{\delta}}} + \frac{13}{15}
+ \ldots
\right).
\ee
Combining both $I_1$ and $I_2$, we obtain the finite term to be 
 -$\frac{5107}{630 \delta}$ after integrating which leads to \eqref{Fgravitino}.

\end{appendix}


\begin{thebibliography}{10}




\bibitem{Weinberg} 
  S.~Weinberg,
  ``Quantum contributions to cosmological correlations,''
  Phys.\ Rev.\ D {\bf 72}, 043514 (2005)
  doi:10.1103/PhysRevD.72.043514
  [hep-th/0506236].

\bibitem{Senatore} 
  L.~Senatore and M.~Zaldarriaga,
  ``On Loops in Inflation,''
  JHEP {\bf 1012}, 008 (2010)
  doi:10.1007/JHEP12(2010)008
  [arXiv:0912.2734 [hep-th]].

\bibitem{Chai} 
  K.~Chaicherdsakul,
  ``Quantum Cosmological Correlations in an Inflating Universe: Can fermion and gauge fields loops give a scale free spectrum?,''
  Phys.\ Rev.\ D {\bf 75}, 063522 (2007)
  doi:10.1103/PhysRevD.75.063522
  [hep-th/0611352].


\bibitem{Tan} 
  H.~S.~Tan,
  ``Noncommutative spacetime geometry and one-loop effects in primordial cosmology,''
  Phys.\ Rev.\ D {\bf 98}, no. 6, 063518 (2018)
  doi:10.1103/PhysRevD.98.063518
  [arXiv:1807.04323 [hep-th]].




\bibitem{Tanaka} 
  T.~Tanaka and Y.~Urakawa,
  ``Loops in inflationary correlation functions,''
  Class.\ Quant.\ Grav.\  {\bf 30}, 233001 (2013)
  doi:10.1088/0264-9381/30/23/233001
  [arXiv:1306.4461 [hep-th]].



\bibitem{Feng} 
  K.~Feng, Y.~F.~Cai and Y.~S.~Piao,
  ``IR Divergence in Inflationary Tensor Perturbations from Fermion Loops,''
  Phys.\ Rev.\ D {\bf 86}, 103515 (2012)
  doi:10.1103/PhysRevD.86.103515
  [arXiv:1207.4405 [hep-th]].




\bibitem{Adshead} 
  P.~Adshead, R.~Easther and E.~A.~Lim,
  ``The 'in-in' Formalism and Cosmological Perturbations,''
  Phys.\ Rev.\ D {\bf 80}, 083521 (2009)
  doi:10.1103/PhysRevD.80.083521
  [arXiv:0904.4207 [hep-th]].




\bibitem{Eugene} 
  P.~Adshead, R.~Easther and E.~A.~Lim,
  ``Cosmology With Many Light Scalar Fields: Stochastic Inflation and Loop Corrections,''
  Phys.\ Rev.\ D {\bf 79}, 063504 (2009)
  doi:10.1103/PhysRevD.79.063504
  [arXiv:0809.4008 [hep-th]].

\bibitem{Mark2017} 
  T.~Markkanen,
  ``Renormalization of the inflationary perturbations revisited,''
  JCAP {\bf 1805}, 001 (2018)
  doi:10.1088/1475-7516/2018/05/001
  [arXiv:1712.02372 [hep-th]].


\bibitem{Mark2013} 
  T.~Markkanen and A.~Tranberg,
  ``A Simple Method for One-Loop Renormalization in Curved Space-Time,''
  JCAP {\bf 1308}, 045 (2013)
  doi:10.1088/1475-7516/2013/08/045
  [arXiv:1303.0180 [hep-th]].




\bibitem{Thaler} 
  Y.~Kahn, D.~A.~Roberts and J.~Thaler,
  ``The goldstone and goldstino of supersymmetric inflation,''
  JHEP {\bf 1510}, 001 (2015)
  doi:10.1007/JHEP10(2015)001
  [arXiv:1504.05958 [hep-th]].

\bibitem{Liam} 
  D.~Baumann and L.~McAllister,
  ``Inflation and String Theory,''
  doi:10.1017/CBO9781316105733
  arXiv:1404.2601 [hep-th].

\bibitem{TASI} 
  D.~Baumann,
  ``Primordial Cosmology,''
  PoS TASI {\bf 2017}, 009 (2018)
  doi:10.22323/1.305.0009
  [arXiv:1807.03098 [hep-th]].


\bibitem{Weinberg:text} 
  S.~Weinberg,
  ``Cosmology,''
  Oxford, UK: Oxford Univ. Pr. (2008) 



\bibitem{WeinbergQFT} 
  S.~Weinberg,
  ``The quantum theory of fields. Vol. 3: Supersymmetry,''
  Cambridge University Press. (2013) 



\bibitem{Moroi} 
  T.~Moroi,
  ``Effects of the gravitino on the inflationary universe,''
  hep-ph/9503210.


\bibitem{Kallosh:1999}
R.~Kallosh, L.~Kofman, A.~D.~Linde and A.~Van Proeyen,
``Gravitino production after inflation,''
Phys. Rev. D \textbf{61}, 103503 (2000)
doi:10.1103/PhysRevD.61.103503
[arXiv:hep-th/9907124 [hep-th]].


\bibitem{Giudice} 
  G.~F.~Giudice, I.~Tkachev and A.~Riotto,
  ``Nonthermal production of dangerous relics in the early universe,''
  JHEP {\bf 9908}, 009 (1999)
  doi:10.1088/1126-6708/1999/08/009
  [hep-ph/9907510].

\bibitem{Bergshoeff:2015}
E.~A.~Bergshoeff, D.~Z.~Freedman, R.~Kallosh and A.~Van Proeyen,
``Pure de Sitter Supergravity,''
Phys. Rev. D \textbf{92}, no.8, 085040 (2015)
doi:10.1103/PhysRevD.93.069901
[arXiv:1507.08264 [hep-th]].

\bibitem{Waldron}
S.~Deser and A.~Waldron,
``(Dis)continuities of massless limits in spin 3/2 mediated interactions and cosmological supergravity,''
Phys. Lett. B \textbf{501}, 134-139 (2001)
doi:10.1016/S0370-2693(01)00109-5
[arXiv:hep-th/0012014 [hep-th]].

\bibitem{Manoukian} 
  E.~B.~Manoukian,
  ``Rarita - Schwinger massless field in covariant and Coulomb-like gauges,
  Mod.\ Phys.\ Lett.\ A {\bf 31}, no. 08, 1650047 (2016).
  doi:10.1142/S0217732316500474

\bibitem{Kosower} 
  D.~A.~Kosower,
  ``Light Cone Recurrence Relations for QCD Amplitudes,''
  Nucl.\ Phys.\ B {\bf 335}, 23 (1990).
  doi:10.1016/0550-3213(90)90167-C


\bibitem{Morrison} 
  D.~Marolf and I.~A.~Morrison,
  ``The IR stability of de Sitter QFT: results at all orders,''
  Phys.\ Rev.\ D {\bf 84}, 044040 (2011)
  doi:10.1103/PhysRevD.84.044040
  [arXiv:1010.5327 [gr-qc]].



\bibitem{Chen} 
  X.~Chen, Y.~Wang and Z.~Z.~Xianyu,
  ``Loop Corrections to Standard Model Fields in Inflation,''
  JHEP {\bf 1608}, 051 (2016)
  doi:10.1007/JHEP08(2016)051
  [arXiv:1604.07841 [hep-th]].

\bibitem{Maldacena} 
  N.~Arkani-Hamed and J.~Maldacena,
  ``Cosmological Collider Physics,''
  arXiv:1503.08043 [hep-th].


\bibitem{Baumann} 
  N.~Arkani-Hamed, D.~Baumann, H.~Lee and G.~L.~Pimentel,
  ``The Cosmological Bootstrap: Inflationary Correlators from Symmetries and Singularities,''
  arXiv:1811.00024 [hep-th].



\bibitem{Senatore2} 
  L.~V.~Delacretaz, V.~Gorbenko and L.~Senatore,
 ``The Supersymmetric Effective Field Theory of Inflation,''
  JHEP {\bf 1703}, 063 (2017)
  doi:10.1007/JHEP03(2017)063
  [arXiv:1610.04227 [hep-th]].


\bibitem{Tan2021}
H.~S.~Tan, ``One-loop corrections to the scalar and tensor spectrum from gravitinos" (Work in progress).



\end{thebibliography}
\end{document}